%% file: main.tex
\documentclass[fleqn,usenatbib]{mnras}

\usepackage{newtxtext,newtxmath}
\usepackage{hyperref}

\usepackage[T1]{fontenc}

\DeclareRobustCommand{\VAN}[3]{#2}
\let\VANthebibliography\thebibliography
\def\thebibliography{\DeclareRobustCommand{\VAN}[3]{##3}\VANthebibliography}
\usepackage{float}

\usepackage{graphicx}	
\usepackage{amsmath}	

\newcommand{\mm}[1]{\mbox{$#1$}} 

\newcommand{\Msol}{\textrm{M}\mm{_\odot}}
\newcommand{\h}{\mm{h^{-1}}}

\newcommand{\sig}{\mm{\sigma_8}}
\newcommand{\bstar}{\mm{\zeta^*}}
\newcommand{\bm}[1]{\boldsymbol{#1}}
\newcommand{\sbr}[1]{_{\textrm{#1}}} 

\newcommand{\degree}{$^{\circ}$}
\newcommand{\kband}{\mm{K}-band}

\title[Calibrating 2M++]%
{Cosmological parameters estimated from peculiar velocity -- density comparisons: calibrating 2M++}

\author[Hollinger and Hudson]{
Amber M. Hollinger$^{1,2}$ 
 and Michael J. Hudson$^{1,2,3}$
\\
$^{1}$Department of Physics and Astronomy, University of Waterloo, 200 University Ave W, Waterloo, ON N2L 3G1, Canada\\
$^{2}$Waterloo Centre for Astrophysics, University of Waterloo, 200 University Ave W, Waterloo, ON N2L 3G1, Canada\\
$^{3}$Perimeter Institute for Theoretical Physics, 31 Caroline St. North, Waterloo, ON N2L 2Y5, Canada
}

\date{Accepted XXX. Received YYY; in original form ZZZ}

\pubyear{2023}

\begin{document}
\label{firstpage}
\pagerange{\pageref{firstpage}--\pageref{lastpage}}
\maketitle

\begin{abstract}
Cosmological parameters can be measured by comparing  peculiar velocities with those predicted from a galaxy density field. Previous work has tested the accuracy of this approach with N-body simulations, but generally on idealised mock galaxy surveys. 
However, systematic biases may arise solely due to survey selection effects such as flux-limited samples, edge-effects and complications due to the obscuration of the Galactic plane. 
In this work, we explore the impact of each of these effects individually as well as collectively using the semi-analytic models from numerical simulations to generate mock catalogues that mimic the 2M++ density field. We find the reconstruction and analysis methods used for our 2M++ mocks produce a value of  $f\sigma_8$ that is biased high by a factor $1.04\pm0.01$ compared to the true value. Moreover, a cosmic volume matching that of 2M++ has a cosmic variance uncertainty in $f\sigma_8$ of $\sim 5\%$. The systematic bias is a function of distance: it is unbiased close to the origin but is biased slightly high for distances in the range 100--180 \h Mpc.  Correcting for this small bias, we find that recent peculiar velocity samples yield $f\sigma_8^{\textrm{lin}} = 0.362\pm0.023$, a value that is in tension with the extrapolations from cosmic microwave background measurements. The predicted peculiar velocities from 2M++ have an error of 170 km s$^{-1}$ that slowly increases with distance, exceeding 200 km s$^{-1}$ only at distances of 180-200 \h Mpc. Finally, the residual bulk flow speeds found in previous work are shown to be not in conflict with those expected in the $\Lambda$ cold dark matter model.
\end{abstract}




\section{Introduction}
Over the past few decades, cosmology has evolved into a precision science. The vast amount of data across a wide range of  redshifts has enabled the precise measurements of cosmological parameters \citep{planck_collaboration_planck_2020, alam_completed_2021,brout_pantheon_2022}.   The standard cosmological model assumes that on all scales gravity is described by general relativity with two additional assumptions; that there is a positive cosmological constant ($\Lambda > 0$) resulting in the Universe's accelerated expansion, and the presence of cold dark matter (CDM). While $\Lambda$CDM has been incredibly successful in explaining observations, both at high and low-redshifts, some  tensions do exist between measured results.   As the uncertainty in these measurements decreases, we are entering an era where even slight deviations from $\Lambda$CDM could potentially show up as a meaningful anomaly in the data. 

There are two fundamental ways to test the standard cosmological model: by studying the Universe's cosmic expansion history over a wide range of redshifts and by measuring the growth rate of cosmic structure. In linear theory, peculiar velocities, which are the deviations in the motions of galaxies from the overall expansion of the universe, constrain the degenerate parameter combination of the logarithmic growth rate $f$ and the amplitude of matter density fluctuations on scales of $8 h^{-1}$ Mpc, \sig. $\Lambda$CDM predicts that the growth rate should be scale independent and related to matter density by $f(z) \approx \Omega_m(z)^{0.55}$ \citep{linder_cosmic_2005}. However, in modified gravity models, it not uncommon for theories to predict growth rates that are scale dependent \citep[as discussed in][]{baker_new_2014}. Peculiar velocities play a crucial role in distinguishing and testing cosmological models, particularly when it comes to fluctuations on the largest scales where the effects of modified gravity become apparent. This makes peculiar velocities an essential tool for disentangling different cosmological scenarios and evaluating the validity of the standard cosmological model \citep{koda_are_2014,howlett_cosmological_2017}. 

One key aspect that makes peculiar velocity measurements particularly important is their ability to map out matter distributions at low redshifts where other observational techniques are limited or less accurate. In particular, there are still tensions related to $S_8 \equiv (\Omega\sbr{m}/0.3)^{0.5}$ and $f \sigma_8$ that typically arise when comparing measurements obtained from different observational probes\footnote{Note that the two parameters are closely related in $\Lambda$CDM: $S_8 = 1.94 \, f\sigma_8 \, (\Omega\sbr{m}/0.3)^{-0.05}$. The last term is negligible for plausible values of the cosmological parameters, varying from $1.0034$ to $0.9968$ as $\Omega\sbr{m}$ is varied from 0.28 to 0.32.}.
Specifically, measurements of cosmic microwave background (CMB) anisotropies by \cite{planck_collaboration_planck_2020} show a tension at the level of $2 -3 \sigma$ when compared to lower redshift probes such as weak gravitational lensing combined with galaxy clustering \citep[eg;][]{ asgari_kids-1000_2021, heymans_kids-1000_2021, abbott_dark_2022, dalal_hyper_2023}

The use of peculiar velocity surveys to measure the growth rate 
and galaxy bias has an extended history \citep[eg. see][]{strauss_density_1995}, and has been done in many different fashions: 
by modelling the density and peculiar velocities fields using a maximum-likelihood approach under the assumption both fields are drawn from multivariate Gaussian distributions \citep{johnson_6df_2014,huterer_testing_2017,howlett_measuring_2017,adams_improving_2017,adams_joint_2020}, 
deriving it from the two-point statistics of direct peculiar velocity estimates for individual galaxies \citep{gorski_cosmological_1989,strauss_density_1995}, via measurements of the galaxy-velocity correlation function \citep{nusser_velocity-density_2017, dupuy_estimation_2019,turner_improving_2021}, and
measuring the momentum power spectrum \citep{park_cosmic_2000,park_power_2006, howlett_redshift-space_2019,qin_redshift-space_2019}. 

The earliest methods of measuring $f$\sig\ focused on predicting the peculiar velocity field using the underlying galaxy density field \citep{pike_cosmological_2005, davis_local_2011, turnbull_cosmic_2012, carrick_cosmological_2015, boruah_cosmic_2020,said_joint_2020, lilow_constrained_2021}. This is the method that we use in our work here.
Linear perturbation theory states that the relationship between matter overdensity, $\delta \equiv \rho / \bar{\rho} -1 $, and peculiar velocities, $\boldsymbol{v}$, is given by  
\begin{equation}\label{eqn:v_r}  \boldsymbol{v}(\boldsymbol{r}) = \frac{H a f(\Omega_m)}{4 \pi}\int \delta(\boldsymbol{r}') \frac{ (\boldsymbol{r}'-\boldsymbol{r})}{|\boldsymbol{r}'-\boldsymbol{r}|^3} d^3\boldsymbol{r}'.
\end{equation}
Formally, this is only true in the linear regime where density perturbations are small, i.e.\ $|\delta| \leq 1$.

However, the total density contrast $\delta$ cannot be measured empirically, as it is dominated by dark matter. Instead, an assumption must be made as to how the observed galaxies trace the underlying total matter. If one assumes linear biasing, the relation is $\delta\sbr{g} = b\sbr{g} \delta$,
where $b\sbr{g}$ is the linear bias and $\delta\sbr{g}$ is the density fluctuation field of the \emph{galaxies}. The same relation is then also true for the standard deviation of fluctuations in an 8 \h Mpc sphere: $\sigma_{8,g} = b_g \sigma_8$. Under this assumption, we can replace $\delta$ in equation (\ref{eqn:v_r}) with $\delta\sbr{g}/b\sbr{g}$ and move $b\sbr{g}$ to the denominator outside the integral.
Comparing the predicted and observed peculiar velocities allows us to measure the degenerate combination $\beta \equiv f/b\sbr{g}$. If linear biasing holds, then by measuring $\sigma\sbr{8,g}$ directly from galaxy surveys, the galaxy bias can be eliminated and we are left with the degenerate cosmological parameter combination 
\begin{equation}
f \sigma_8 = \beta \sigma_{8,g}\,.
\end{equation}

Another motivation for understanding the accuracy and precision of predicted peculiar velocities arises from their use as corrections to the observed redshift and the impact of these on measurements of the Hubble parameter and other cosmological parameters measured from the Hubble diagram. In the case of the Hubble parameter, early work focused on the effect of Virgo infall peculiar velocities. For more recent supernova-based calibrations, \cite{neill_peculiar_2007} showed that peculiar velocity corrections impacted the fit to the Hubble diagram and hence the measurements of the cosmological parameters. More recently, the peculiar velocity predictions of \citet[hereafter C15]{carrick_cosmological_2015} have been absorbed into the redshifts of the supernova rung of the distance ladder when measuring $H_0$ \citep{riess_24_2016, riess_comprehensive_2022, peterson_pantheon_2022}. Therefore, the uncertainties on these corrections make up part of the total error budget in $H_0$.

Velocity-velocity comparisons, while a valuable method to probe cosmological parameters, may be subject to several potential biases that need to be carefully considered in their interpretation. 
Redshift data from comprehensive compilations like 2M++ \citep{lavaux_2m_2011} allow for the reconstruction of density and velocity fields to be performed more accurately than ever before. Our previous work, \citet[hereafter HH21]{hollinger_assessing_2021},  investigated how the choice of density field tracer affected the recovered values of $\beta$ and $f\sigma_8$. We measured the dependence of these quantities on the choice of tracer object used to create the density field: dark matter particles and haloes from N-body simulations, or galaxy observables, such as stellar mass and luminosity from the semi-analytic models (SAMs). That work did not explore in detail any biases that arise due to survey selection effects. 

In this work, we explore the additional uncertainties that arise due to survey selection effects in the 2M++ survey compilation, and how these selection effects impact measurements of the cosmological parameters. We will focus on weighting the density field primarily by the `observed' \kband\ luminosity in the presence of a flux-limited redshift survey as was done in C15. 
The lack of redshift data in obscured regions such as the Zone of Avoidance (hereafter ZoA) result in missing large-scale structures, which also impacts inferred parameters such as $\beta$.

We approach this work in the same vein as HH21, where instead of simulating all the observational properties of the surveys simultaneously, our approach focuses on considering the different physical effects that can introduce bias or uncertainty into our results one-by-one. This work is similar in spirit to that of \cite{lilow_constrained_2021}, who studied similar biases and correction procedures for cosmological parameters based on the reconstructed 2MRS density field, which is slightly shallower than 2M++.

This paper is organised as follows. 
We present the method used to generate our mock 2M++-like surveys in Section \ref{sec:gen$K$-band}. Section \ref{sec:weights} focuses on the effects generated solely due to flux limiting the data. In Section \ref{sec:ZoA} we explore how the treatment of the Zone of Avoidance impacts measurements. Section \ref{sec:combined} investigates how these two biases are impacted when dealing with a full mock realisation of 2M++. We present how these results impact our previous measurements in Section \ref{sec:impacts}. Finally, we present our conclusions in Section \ref{sec:discussion}.

\section{Generating mock 2M++ surveys and density fields}\label{sec:gen$K$-band}

In order to study peculiar velocities sourced by a density field on large scales (and since the integral in equation (\ref{eqn:v_r}) is over all space), one requires a galaxy catalogue to not only be as deep as possible, but also to cover a large fraction of the sky. One such catalogue that meets the latter criterion is  the 2M++ galaxy redshift compilation \citep{lavaux_2m_2011}. The redshifts of 2M++ were obtained from three redshift surveys with different depths and sky coverage: the Two-Micron All-Sky Redshift Survey \cite[2MRS; ][]{huchra_2mass_2012}, the third data release Six-degree Field Galaxy Survey \citep[6dFGS DR3; ][]{jones_6df_2009}, and the Sloan Digital Sky Survey Data Release 7 \citep[SDSS DR7; ][]{abazajian_seventh_2009}. 
The resulting catalogue is highly complete to K $\leq 12.5$ for regions covered by either 6dFGRS or SDSS, otherwise for the regions covered only by 2MRS is complete to K $\leq 11.5$. In C15, a limiting distance of 200 \h Mpc was adopted for the region with K $<12.5$. At this depth, the luminosity weights are typically a factor of 6 or less. The  K $<11.5$ region is a magnitude shallower, or a factor $10^{-0.2}=0.63$ in distance, so a limit of 125 \h Mpc was adopted. For consistency with C15, we adopt the same distance limits here.

\subsection{Simulation Data}\label{sec:sim_data}

We used the MultiDark Planck 2 simulation (MDPL2), which is part of MultiDark project, a suite of dark matter only cosmological simulations \citep{klypin_multidark_2016}. The simulations all assume a flat $\Lambda$CDM cosmology with cosmological parameters: $\Omega_\Lambda=0.692885$, $\Omega_M=0.307115$, $h=0.6777$,  $\sigma\sbr{8,linear}=0.8228$, and $n_s=0.96$, such that they are consistent with Planck results \citep{planck_collaboration_planck_2020-1}. The (non-linear) $\sigma\sbr{8,m}$, measured from the particles, is\footnote{For reasons of computational efficiency, in HH21, $\sigma\sbr{8,m}=0.95$ was calculated using particle counts in small ($\sim 1$ Mpc) voxels that were themselves within an 8 \h Mpc radius sphere. A careful calculation shows that the actual volume of the voxels corresponds to that of a sphere of radius 7.77 \h Mpc.} 0.925.  MDPL2 uses 3840$^3$ dark matter particles with masses equal to $1.51\times 10^9$ \h \Msol. Rockstar and ConstistentTrees were used to identify more that 10$^8$ haloes \citep{behroozi_rockstar_2012}, and generate  halo merger trees \citep{behroozi_gravitationally_2013}. MDPL2 has a periodic box of length 1000 $h^{-1}$ Mpc and is evolved from a redshift of 120 to 0, along with a varying physical force resolution level from 13-5 \h\ kpc and various implemented physics.  This work uses the snapshots taken at $z=0$. Both the halo and galaxy catalogues were obtained from the COSMOSIM database \footnote{\url{www.cosmosim.org}}. 

The  SAG \citep{cora_semi-analytic_2018} and SAGE \citep{croton_semi-analytic_2016} SAMs were calibrated to generate galaxy catalogues using the MDPL2 simulation. These models include the most relevant physical processes in galaxy formation and evolution, including: galaxy mergers, radiative cooling, chemical  enrichment, supernova feedback and winds, star formation, starbursts, and disc instabilities.  A comprehensive review of the models can be found in \cite{knebe_multidark-galaxies_2018}. 
The presence of baryons in full hydrodynamic simulations does introduce some effects on their host dark matter haloes \citep[e.g.][]{schaller_baryon_2015,velliscig_intrinsic_2015}. This impacts the mass of haloes and their mass distribution on small scales which alters their substructure as well as clustering measurements \citep{vaandaalen_impact_2014}. As these effects are limited to scales of $k \gtrsim 0.5 h$ Mpc$^{-1}$ \citep{hellwing_effect_2016}, we expect the impact on peculiar velocities to be small since they depend on large-scale structure. To test this, we have compared the predicted peculiar velocities for two different assumed power spectra: a simple non-linear power spectrum based on halofit \citep{mead_accurate_2016}, and a modified non-linear power spectrum with baryonic feedback \citep{mead_hmcode_2021} from high-temperature AGN ($\log10(T_{\rm AGN}/{\rm K}) = 8.5$). These two scenarios differ by less than 0.7\% in the predicted peculiar velocity speed. Hence we ignore this small effect in the remainder of this work.



From the MDPL2 survey, we generate 15 independent (non-overlapping) finite volume realisations of radius 200 \h Mpc to be consistent with the hard radius cuts applied in the analysis of C15. Any galaxy that exists outside of each mock local Universe is ignored.  For each sphere, we take the Cartesian velocities and project them to a radial velocity. 

\subsection{Generating \kband\ Luminosities}
\begin{figure}
    \centering
    \includegraphics[width=.5\textwidth]{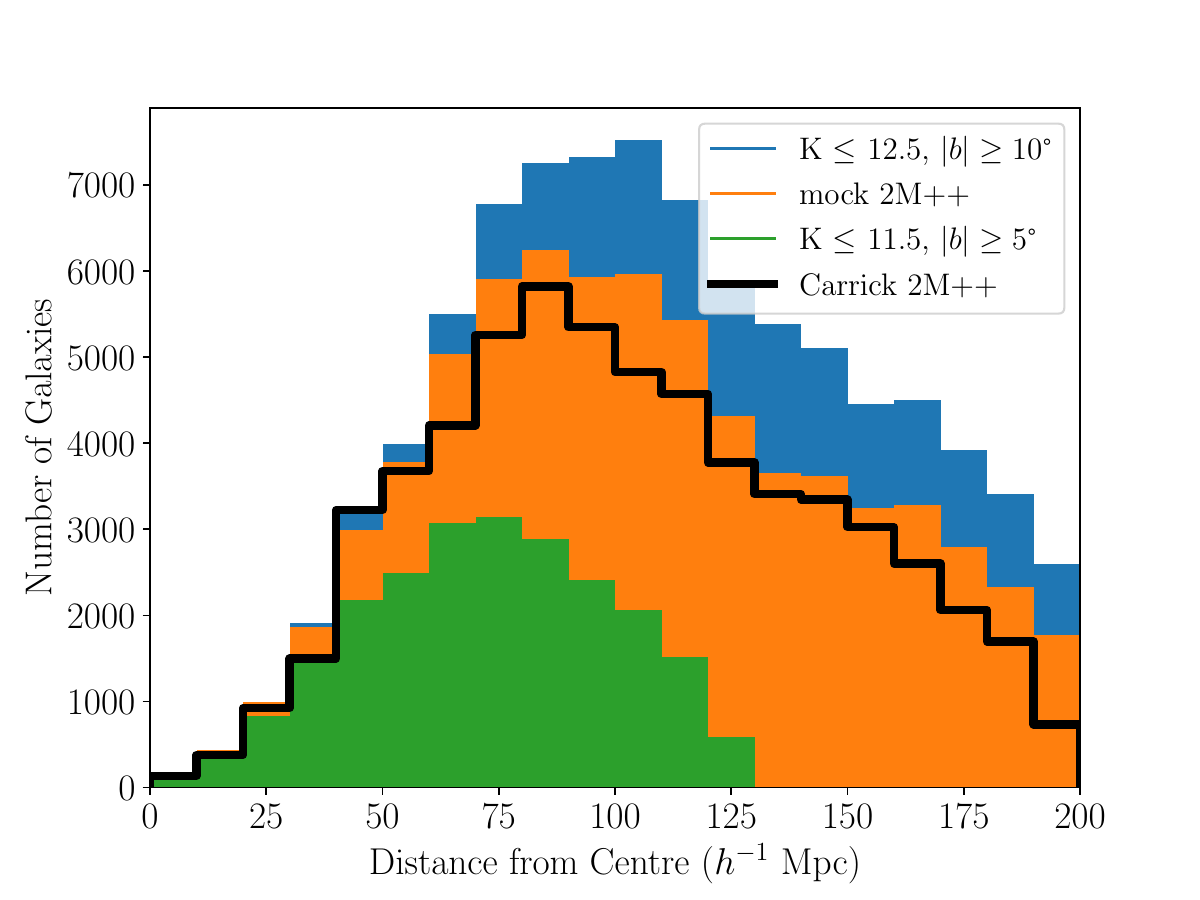}
    \caption{Histogram of SAG galaxies in a mock local Universe that could have been observed given the 2M++ survey magnitude limits, as a function of comoving distance. The black line represents the 2M++ galaxy data observed, taken from C15. The blue and green histograms represent the two \kband\ magnitude flux limits of 11.5 and 12.5.
    The former ignores data that is within $\pm$ 5\textdegree\ from the Galactic plane (matching the 2MRS survey) and the latter $\pm$ 10\textdegree\ (as in the 6dFGRS and SDSS redshift surveys).
    The orange histogram shows the mock Universe with the 2M++ survey mask and conditions applied.}
    \label{fig:n(z)}
\end{figure}

The cosmological simulation and SAMs used in this work provide both galaxy halo and stellar masses. However,  as we are interested in generating realistic 2M++ mock catalogues of the local Universe,  we need the mock galaxy observables to be consistent with this survey,  which in this case uses \kband\ luminosities. However, as this observable is not provided by the SAMs, in this work, \kband\ luminosities were generated using the abundance matching technique.  This provides a simple and effective method which traditionally is used to associate galaxies with dark matter halos \citep{marinoni_mass--light_2002, kravtsov_dark_2004, behroozi_comprehensive_2010, moster_constraints_2010}.  As \kband\ luminosity is a good proxy for stellar mass, the inverse should also be true. A simple explanation of this technique is that given a stellar mass function and a galaxy luminosity function, \kband\ luminosities are assigned such that the most massive galaxy (according to its stellar mass) is assigned the brightest \kband\ luminosity, the second most massive galaxy has the second brightest luminosity, and so on. 

In the real Universe, there may be additional scatter in the mapping between stellar mass and \kband\ luminosity that is not captured by our prescription. However, our method does capture the dominant source of scatter, namely the scatter in the stellar mass to halo mass relation. In HH21, we found that this scatter was much larger in the SAGE SAM than in the SAG SAM (see HH21 fig. 6): 0.39 dex and 0.15 dex, respectively, for halos in the range $10^11 - 10^13$ $\h M_\odot$.  In spite of this, the fitted values of \bstar for these two SAMs differ by less than 1\% (see Fig. \ref{fig:SAGEvSAG}). Therefore we conclude that scatter does not significantly bias the fitted \bstar.


To replicate the \kband\ luminosity function of the 2M++ survey, we use the Schechter function \citep{schechter_analytic_1976} parameters values associated with the survey to do the abundance matching: 
\begin{equation}
    \Phi(L)=\left(\frac{\phi_s}{L_*} \right) \left(\frac{L}{L_*} \right)^{\alpha} \exp \left( \frac{-L}{L_*}\right) ,
\end{equation}
where $\phi_{*}$  is the density normalisation, $L_{*}$ is the absolute magnitude break in the luminosity, and $\alpha$ is a power-law parameter to be determined. For 2M++, $\alpha = 0.73$, $L^*=-23.17 - 5\log_{10} h$ and $\phi^*=1.11 \times 10^{-2} h^3$ Mpc$^{-3}$  \citep{lavaux_2m_2011}. More generally, this approach is likely complicated by scatter in the galaxy stellar mass-luminosity relation \citep{tasitsiomi_modeling_2004, behroozi_comprehensive_2010}, but for the purpose of this paper we ignore this additional complication as there is already scatter present in the halo stellar-mass relation of MDPL2 \cite[see][for more details]{hollinger_assessing_2021}.

Shown in Fig.\ \ref{fig:n(z)} are redshift histograms from one of the mock volumes used in this paper. It demonstrates the count of galaxies as a function of distance after luminosity and distance cuts have been applied, see Section \ref{sec:magcuts-weights} in which these are described. 
Specifically, we show two flux-limited surveys (6dFGRS/SDSS and 2MRS assuming full celestial sphere coverage) and a mock that matches the variable depth of the 2M++ compilation. The latter is a good match to the observed redshift distribution in 2M++ (from C15).

\subsection{Apparent Magnitude Limits and Weights}\label{sec:applying}

A complication of analyses performed with galaxy redshift surveys is that the data used to generate the density field are typically flux-limited. In this paper, we are interested in replicating the 2M++ survey, as described in \cite{lavaux_2m_2011}, to understand the different biases associated with the survey. 
To be able to match the survey conditions, we must also apply the relevant flux limits to the mock data. As we are mainly interested in how the various effects individually and collectively impact our measurements of cosmological parameters, we apply the 11.5 and 12.5 apparent magnitude limits of the survey independently, to see how each impact the measurements. In order to do so, however, we need first to account for survey incompleteness by weighting the `observed' objects, before constructing a density field, using the weighting schemes described in C15. 

Here we summarise the method used to calculate the weights applied to galaxies that would be observed in our mock Universe catalogues.
To determine whether a galaxy would be visible in our mock catalogues, we calculate the minimum luminosity each galaxy would have to be at their given distance from the centre of our local  mock Universes, in order to be be visible for the given apparent magnitude limits of 2M++. Any galaxy whose luminosity falls beneath this minimum is excluded from the calculations of the density field.

For each mock, we generate appropriate weights to be applied to individual galaxies to account for survey incompleteness in the outer regions, using the luminosity function.  
There are multiple ways these weights can be generated, however our prescription to account for incompleteness is done by weighting galaxies using a method similar to \cite{davis_survey_1982}. Under the assumption that the relation between the galaxy and underlying dark matter density fields is linear, we can use both the galaxy number and luminosity density fields as a proxy to construct $\delta_g$. In this paper we will explore both the number and luminosity weighted schemes, however our analysis will focus mainly on the latter, unless otherwise specified. 

For the number-density field case, we weight our `observed' galaxies by the number of galaxies at a given distance that are not observed due to survey magnitude limits. In this scheme galaxies are  weighted by:
\begin{equation}
    w^N(r) = \frac{N\sbr{average}}{N\sbr{observed}(r)} = \frac{\int^{\infty}_{L\sbr{min}} \Phi (L) dL}{\int^{\infty}_{4 \pi r^2 f\sbr{min}} \Phi (L) dL} .
\end{equation}

The \kband\ luminosity-density should be a good proxy for stellar-mass-weighted density field, which in turn may be a better proxy for the density of dark matter than simply weighting by galaxy number.
For this luminosity-weighted case the weights are determined via: 
\begin{equation}
    w^L(r) = \frac{L\sbr{average}}{L\sbr{observed}(r)} = \frac{\int^{\infty}_{L\sbr{min}} L \Phi (L) dL}{\int^{\infty}_{4 \pi r^2 f\sbr{min}}L \Phi (L) dL} .
\end{equation}

\section{Recovery of the Cosmological Parameters}\label{sec:weights}

\begin{figure*}
    \centering
    \includegraphics[width=\textwidth]{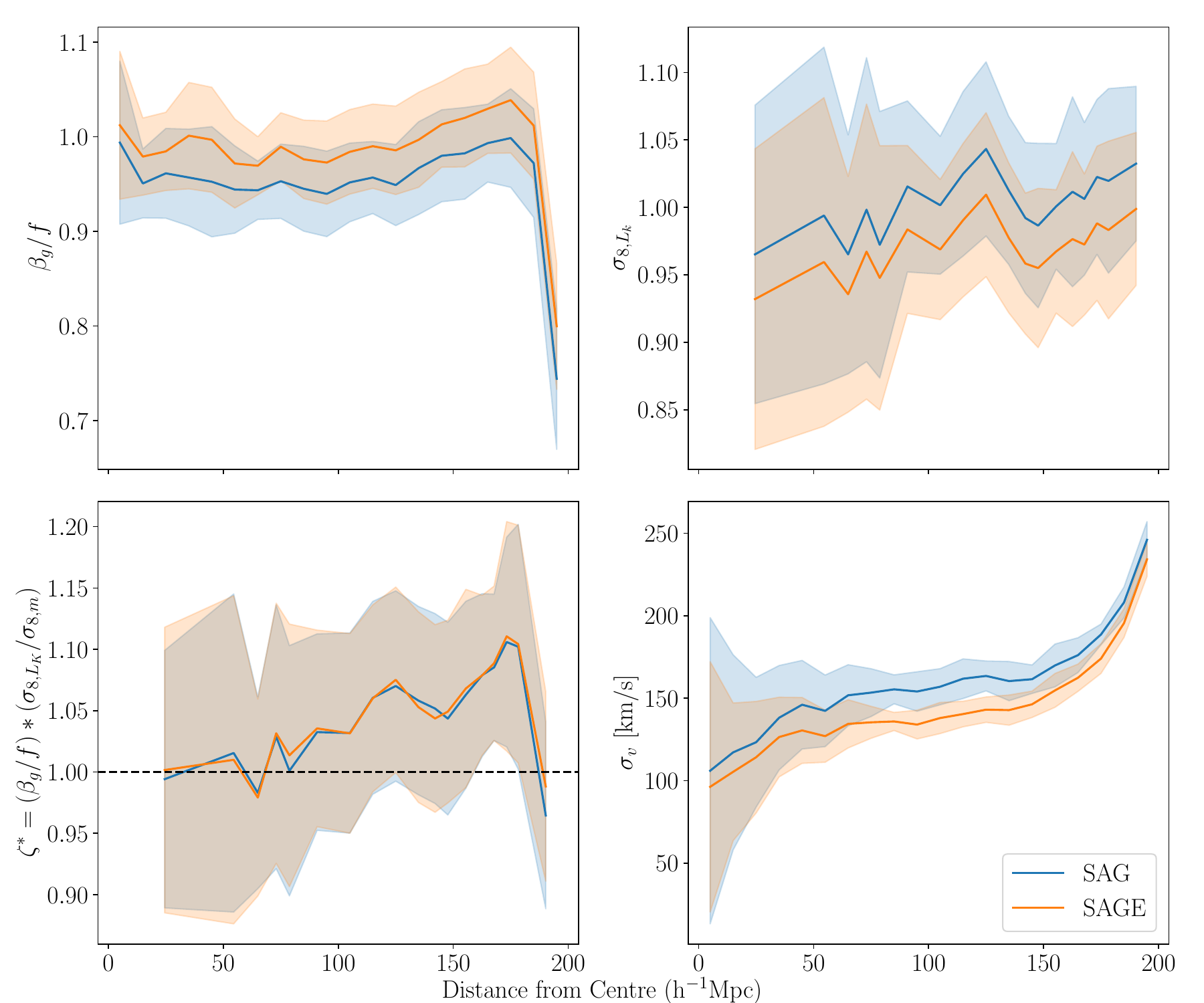}
    \caption{Velocity field parameters from the volume-limited luminosity-weighted  mock density fields, obtained from using SAG (blue) or SAGE (orange) stellar masses to generate $k$-band luminosities (see text for details). In each panel, the solid line represents the average parameter values of mocks and the lighter semi-transparent band represents the standard deviation (from mock to mock, i.e.\ cosmic variance) as a function of radius from the centre of the mock local Universes.
    Top Left panel:  Measured value of $\beta/f$ in shells of 10 \h Mpc out to imposed survey limits.
    Top Right panel: Measurement of \sig\ performed in equal volume bins, so that roughly the same number of \sig\ spheres are located within each shell. 
    Bottom Left Panel: Measured value of $\bstar=(\beta/f) (\sig / \sigma\sbr{m})$, as we correct for both the $f$ parameter and the true \sig\ measured from the global particle value of 0.925, a value of \bstar of unity, shown by the black dash line, indicates where this velocity--density cross-correlation method is unbiased. 
    Bottom Right Panel: The scatter between the measured $N$-body and predicted peculiar velocity associated with each of the linear regression slopes, with the same binning as the top left panel.
    Note that the binning in the panels are slightly inconsistent: this is due to the fact that there are a limited number of non-overlapping 8 $h^{-1}$ Mpc spheres that can be fit, as a result the values that are dependent on \sig\ are binned such that the bins have approximately equal number of spheres.}\label{fig:SAGEvSAG}
\end{figure*}

Having calculated the density field according to some tracer population of galaxies (e.g., luminosity-weighted or number-weighted), we now turn to assess how well these density fields can be used to predict the N-body peculiar velocities and whether there is bias in cosmological parameters, such as $f \sigma_8$. In Section \ref{sec:method}, we describe how the fits are performed, while Section \ref{sec:vollim} discusses the results for the simple case of a volume-limited sample, and Section \ref{sec:magcuts-weights} presents the results for the more realistic flux-limited case.

\subsection{Comparison of predicted and actual peculiar velocities and fitted cosmological parameters}\label{sec:method}

The method used to estimate peculiar velocities from the mock catalogues, and compare these to the actual halo peculiar velocities in the simulations, is designed to follow as closely as possible what is done in comparisons between the reconstructed 2M++ predictions and observed peculiar velocity samples \citep[C15,][]{boruah_cosmic_2020, said_joint_2020}.

\subsubsection{The density and peculiar velocity fields}
Our procedure for generating the density and predicted peculiar velocity fields can be summarised as follows:
\begin{enumerate}

\item The density field is obtained by taking the tracers (generically denoted with subscript $t$ representing, galaxy luminosity, number or halo mass) and placing them onto a padded three-dimensional grid.  In the flux-limited cases, this includes weighting the galaxies (see Section \ref{sec:applying}). The density field is composed of every object that could be observed given the relevant flux limits, so both objects classified as central and satellites are used. 

Then, based on the density of tracers, we calculate the density perturbation field, $\delta\sbr{t}(\boldsymbol{r})$, and smooth it in Fourier space using a Gaussian smoothing kernel of radius 4 \h Mpc. This smoothing scale is chosen based on the values used in C15 and HH21.

\item The velocity field is obtained from the smoothed density perturbation field via the Fourier-transformed version of equation (\ref{eqn:v_r}),
\begin{equation}
\bm{v_k}=i H a f \frac{\bm{k}}{|\bm{k}|^2}\delta_k\,.
\end{equation}
This yields the predicted peculiar velocity field in Cartesian coordinates. 
Note that, in practice, the fields are on a three-dimensional grid in units of $h^{-1}$ Mpc. Therefore, $\bm{k}$ is in units of $h$ Mpc$^{-1}$ and so writing $H_0 = 100 h$ km s$^{-1}$ Mpc$^{-1}$, one sees that all factors of $h$ cancel and the predictions are independent of the assumed value of the Hubble parameter. 
\item  By projecting along the line of sight the Cartesian velocity fields are then used to generate a radial peculiar velocity field.  
\end{enumerate}

\subsubsection{Cosmological parameters from the comparison of observed and predicted peculiar velocities}

Having calculated a predicted peculiar velocity field, we now wish to compare these to the true N-body peculiar velocities of halos to assess any level of systematic bias in the recovered cosmological parameters. Observationally, these parameters are found by regressing, or fitting via maximum likelihood, the measured peculiar velocities against the predictions. Our procedure for fitting the cosmological parameters from the mocks is as follows.
\begin{enumerate}
\item 
We first predict the peculiar velocities by integrating over the tracer density field as described above.
To account for potential contributions from external sources outside the volume, we include an ``external'' bulk flow, $\boldsymbol{V}\sbr{ext}$, in our analysis. The presence or absence of this bulk flow provides further insights into cosmic structures beyond what is captured by local observations alone, as discussed in Section \ref{sec:bulkflow}.
In summary, we modify equation (\ref{eqn:v_r}) so that it reads: 
\begin{equation}
\boldsymbol{v}\sbr{pred}(\boldsymbol{r}) = \frac{H a f(\Omega_m)}{4 \pi}\int_{\textrm{survey}} \delta\sbr{t} (\boldsymbol{r}') \frac{ (\boldsymbol{r}'-\boldsymbol{r})}{|\boldsymbol{r}'-\boldsymbol{r}|^3} d^3\boldsymbol{r}'
+ \boldsymbol{V}\sbr{ext}\,.
\label{eqn:v_r_density}
\end{equation}

\item The true N-body radial peculiar velocities are regressed on the predicted radial peculiar velocities. The slope of the fit then gives a measurement of $\beta\sbr{t}/f = 1/b\sbr{t}$, as well as $\boldsymbol{V}\sbr{ext}$.

\item There is a small correction that needs to be made to the measured values of $\beta\sbr{t}/f$, to account for differences between what is done in this paper versus what has been done with the real 2M++ density field in C15. The galaxies in the 2M++ survey have redshifts but do not have precise distances. However, the equations used in calculating the velocities require the latter. The process used in C15 of determining the distances from the redshifts is referred to as ``reconstruction''. They used an iterative procedure to accomplish the reconstruction based on \cite{yahil_redshift_1991} and the grouping method of \cite{huchra_groups_1982}. This process results in the reconstructed field having a slightly higher $\beta$ (a factor $1.03$ at 4 \h Mpc smoothing) than if the true distances are used (see discussion in C15, Appendix A). As a result, all measurements of $\beta$ in this work have been increased by this factor to simulate what would have been found had we reconstructed the mock density fields. 

\item We calculate $\sigma\sbr{8,t} $ of the mock density field by placing non-overlapping spheres over the entirety of the grid and ignore any sphere that includes areas that either falls outside of the survey limits or is partially located within the ZoA.

\item We then multiply these two quantities together and divide by the known particle $\sigma\sbr{8,m}$ to obtain 
\begin{equation}
\bstar \equiv \left(\frac{\beta}{f}\right) \left(\frac{\sigma\sbr{8,g}}{\sigma\sbr{8,m}}\right)\ .
\end{equation} 
If the method is an unbiased estimator of $f \sigma_8$, we expect $\bstar = 1$\footnote{Note that this parameter is equivalent to that defined as $\beta^*$ in HH21, we change our  notation here for clarity.}, but if there is a bias this allows us to calibrate it.

\item We also calculate the velocity scatter around the regression, $\sigma_v$. This is a measure of how well our mock 2M++ density fields can predict the peculiar velocities. 

\end{enumerate}

In observational studies of the distance scale and peculiar velocities, it has been standard practice to combine the measured distances and redshifts of individual galaxies that are in groups or clusters \citep[e.g.,][]{sandage_steps_1975, de_vaucouleurs_extragalactic_1979, aaronson_infrared_1979, lynden-bell_photometry_1988, tully_cosmicflows-4_2023}. This is because the motions of individual satellite galaxies, with respect to the centre of mass of the galaxy group, are essentially random or ``thermal'' noise, and not described by equation (\ref{eqn:v_r}).  In C15, this was also done, using the ``friends-of-friends'' groups defined by \cite{lavaux_2m_2011}  and taking the mean redshifts of the grouped galaxies when constructing the density field. C15 also used the group redshifts of Tully-Fisher peculiar velocity galaxies where appropriate. Finally, the \cite{turnbull_cosmic_2012} `A1' supernovae sample used by C15 also used the mean cluster redshift for supernovae appearing in galaxy clusters. Therefore, for consistency, here we compare the predicted peculiar velocities with the peculiar velocities of the halo as a whole, or, rather, its proxy in MDPL2, its central galaxy. In the SAG and SAGE SAMs, by construction the central galaxies have the same redshifts as their host ``friends-of-friends'' halo.

\subsection{Volume-Limited Samples}\label{sec:vollim}

Before considering the more realistic case of flux-limited samples, it is interesting to consider volume-limited samples as these make the biases close to the survey edges clear. For the analysis performed in this section, the density field was populated with the $K$-band luminosities of all galaxies located within the survey limits of 200 \h Mpc. We note that regardless of what density tracer is used, only objects that could be observed given the flux limits of 2M++ are used as velocity tracers.

Fig.\ \ref{fig:SAGEvSAG} shows recovered parameters for a volume-limited mock galaxy sample: the mock luminosities generated from the either the SAG or SAGE models are used, with no apparent magnitude cuts applied, and with a minimal stellar mass lower limit of $10^{9}$ $M_{\odot}$ applied to the input galaxy catalogue. As demonstrated in the top left panel, the recovered value of $\beta/f$ (which should be proportional to the inverse of the galaxy bias) is strongly influenced by the effects of the survey edges. Within 20 \h Mpc of the survey edges, the recovered $\beta$ is almost 30\% smaller than at other radii. This makes sense if one considers a galaxy at the survey edge: the density field beyond the outer edge, for example, is set to $\delta\sbr{g} = 0$, so, for that galaxy, the contribution to the integral in equation (\ref{eqn:v_r}) is from approximately half of the sky. The predicted peculiar velocities there will therefore be noisier. Since we are regressing the N-body peculiar velocities on those predicted, this additional noise in the independent variable will flatten the fitted slope, leading to a lower value of $\beta/f$. This increase in scatter in the predicted velocities, shown in the bottom right panel, is indeed highest near the survey edges where velocity fields are most heavily influenced by the lack of data beyond the edge. Measurements of \sig\ (top right panel) are unaffected by edge effects and remain fairly constant across radial shell measurements.

Fig.\ \ref{fig:SAGEvSAG} also shows the $1\sigma$ range of the recovered parameters from mock-to-mock. This scatter is primarily driven by cosmic variance. As described in more detail in HH21, each mock 2M++ region, which typically has an effective volume of a 175 \h Mpc sphere, may be under- or over-dense (which affects the calculation of $\delta$, and hence $\beta$), or may have variance in the locally-measured \sig. These ultimately drive cosmic variance in \bstar.

While the global values of $\beta$ and \sig\ are respectively higher and lower in SAG compared to SAGE, we find that measurements of \bstar\ for both are effectively identical with comparable uncertainties. We note, however, that the SAG model does consistently generate slightly more scatter ($\sim$ 20 km s$^{-1}$) in the measured $\sigma_v$. For remainder of this paper, we shall focus on the \kband\ luminosities generated from the SAG SAM unless otherwise specified.  

\subsection{Effects of Magnitude Cuts and Weights}\label{sec:magcuts-weights}
\begin{figure*}
    \centering 
    \includegraphics[width=1.1\textwidth]{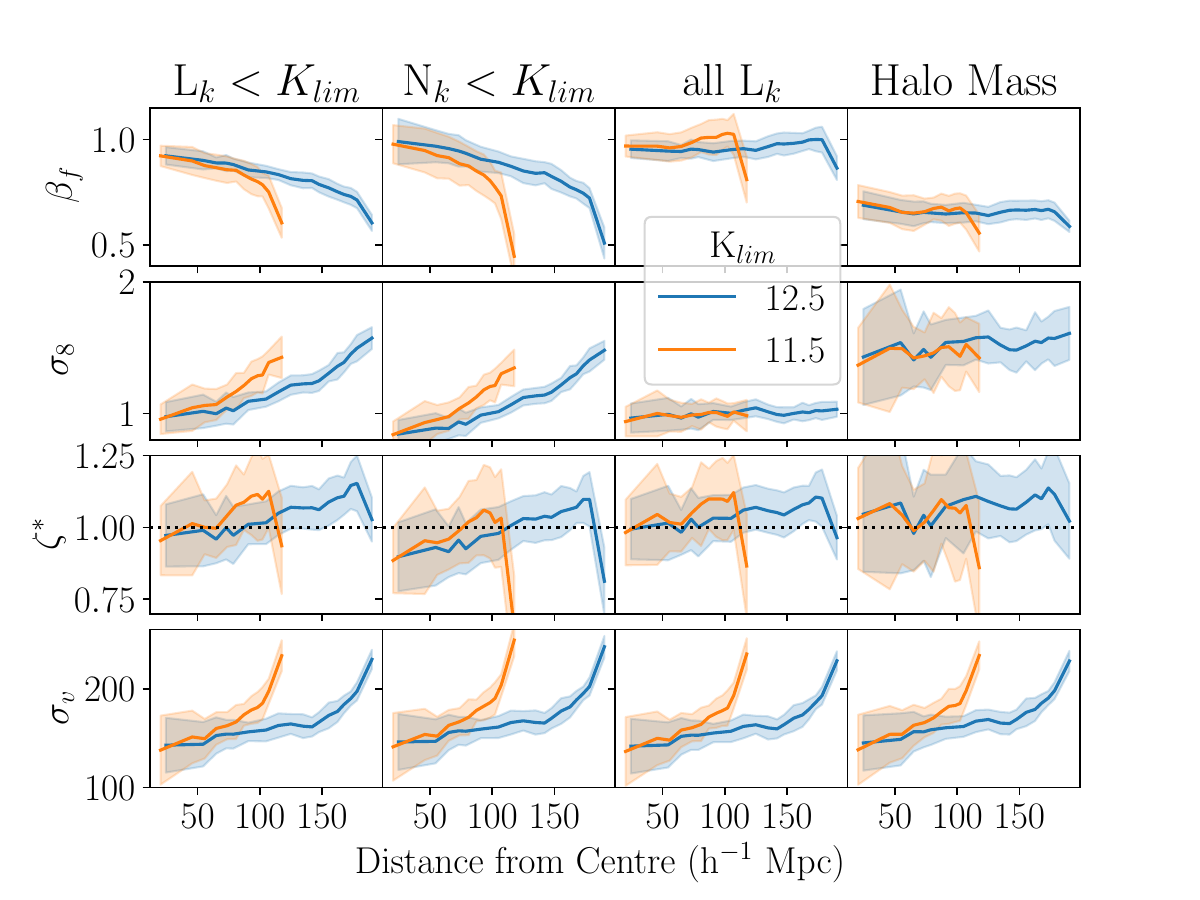}
    \caption{Comparison of differently-weighted density fields and the effect on recovered values. The blue and orange lines represent the two imposed survey limits for 2M++ corresponding to 200 and 125 \h Mpc, respectively, while the coloured bands have the same meaning as in Fig. \ref{fig:SAGEvSAG}. 
    The rows show the same measured quantities as in Fig. \ref{fig:SAGEvSAG},  as a function of comoving distance from the centre.  Each column represents the various tracers/methods used to weight the density weighted field and predict the peculiar velocities. The weightings of the density field are: (1,3) galaxies' \kband\ luminosities, (2) number of galaxies, and (4) halo mass. The data in the first two columns have flux limits and weights applied, corresponding to the \kband\ apparent magnitude limits of 12.5 and 11.5. The velocity tracers in all cases remain the same.}
    \label{fig:12.5v11.5}
\end{figure*}

We now turn to the more realistic case, where the redshift survey is flux-limited. In Fig.\ \ref{fig:12.5v11.5}, we show how our flux-limited density fields (with appropriate weights) compare to the volume-limited case, as well as to the dark matter halo field directly. This is done for both the \kband\ magnitude limits in 2M++ (12.5 and 11.5) with hard distance limits of 200 and 125 $\h$ Mpc, respectively. Each column represents a different density field tracer. From left to right these are: the flux-limited \kband\ luminosities, the flux-limited number of galaxies, the volume-limited \kband\ luminosities and the halo masses. The first row shows the fitted value of $\beta/f$ as a function of distance. Regardless of tracer used in the density field, all are impacted by edge effects close to the survey edges. The fitted $\beta/f$ decreases as a function of distance from the centre of the sphere for the flux limited cases, while remaining constant in the volume-limited cases as expected. This is likely due to a combination of two effects. First, at larger distances the galaxies are more luminous and hence have a higher bias, driving down $\beta$. Second, the field becomes noisier and so the fit will tend to flatten (lower) $\beta$. This same effect led to a drop in $\beta$ near the survey edges, as discussed in the previous subsection.   Similarly, the second row shows that  $\sig$ increases due to the increasing sparsity in the `observed' number of galaxies as a function of distance and the increase in luminosity of those selected, hence an increase in bias. However, in all cases the cases the average recovered value of $\bstar$, shown in the third row, remains fairly constant as a function of radius, and is close to unbiased, although still subject to cosmic variance. The number-weighted case performs less well than the luminosity-weighted case, as expected, since these weights are higher at large distances, which leads to noisier density and peculiar velocity fields. Both the volume-limited cases have consistent values of $\beta$ and $\sig$, this is reflected in the recovered value of $\bstar$. However, the halo-mass-weighted case does show the most variation in its measurement, due to the higher level of fluctuation in the halo-mass-weighted $\sig$. Finally, the fourth row shows the scatter in our velocity predictions. As before, the scatter increases rapidly for locations near the survey edges. We note this effect is strongest for the number weighted case. We find that the velocity scatter $\sigma\sbr{v}$ in the inner regions ($\sim 150$ km s$^{-1}$) is comparable to what was found in our previous work (HH21), which focused on volume-limited samples.

The global (excluding the outer 20 \h Mpc)  measured values of $\bstar$ and $\sigma_v$ for this analysis can be found in Table \ref{table:Tab1}. These are weighted by the number of velocity tracers at each distance in a 2M++-like flux-limited sample. As before, SAG produces slightly poorer predictions than those based on the SAGE catalogues. The SAGE velocity scatter is consistent with that assumed by C15 ($\approx 150 km s^{-1}$), while SAG's scatter is on the order of 20 km s$^{-1}$ higher.

\section{Complications due to the Zone of Avoidance}\label{sec:ZoA}

\begin{figure*}
    \centering
    \includegraphics[width=\textwidth]{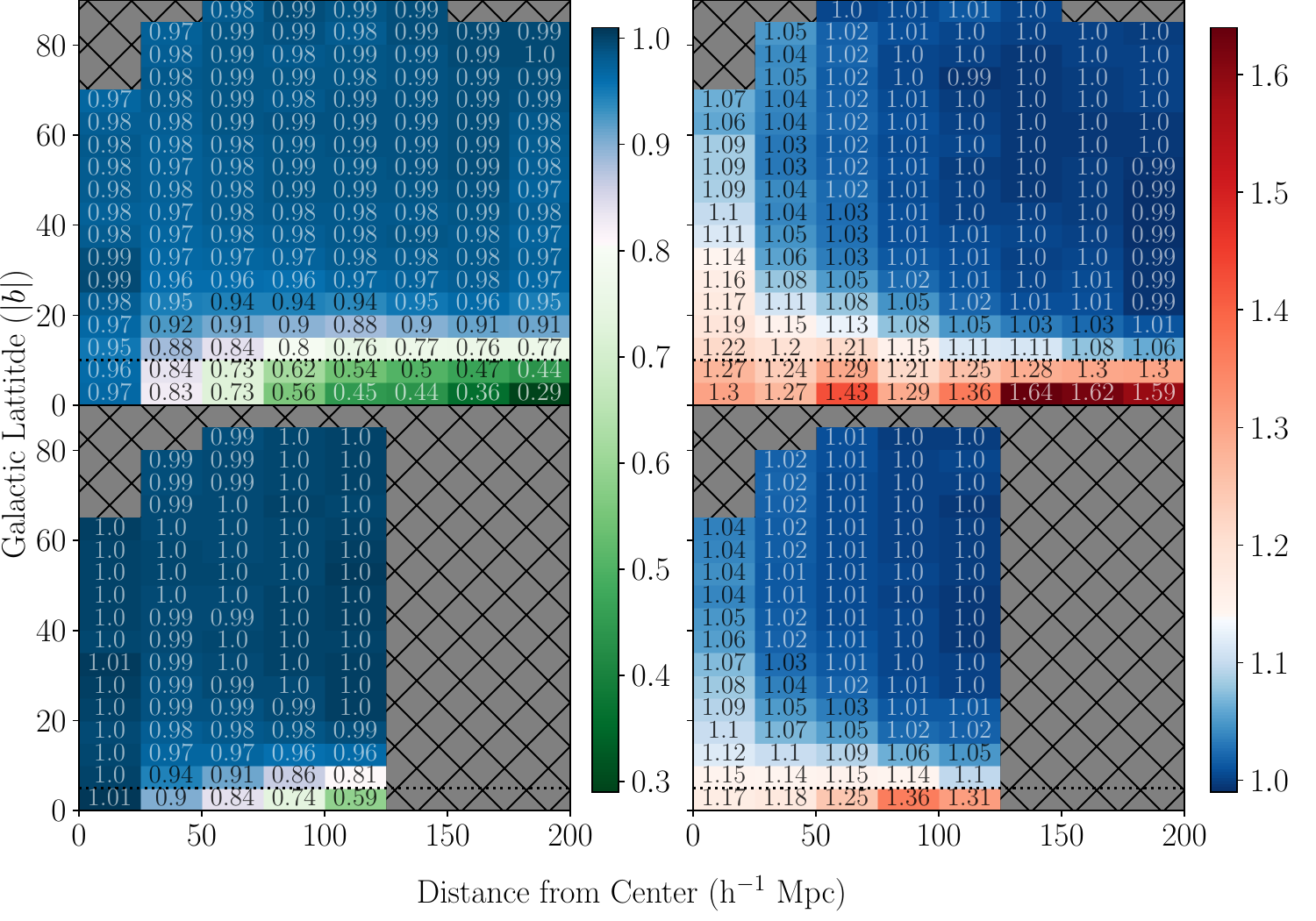}
    \caption{This plot demonstrates ratio of the fitted $\beta/f$  measurements, when using either the cloning (left column) or uniform (right column) prescription for the ZoA, over the fitted $\beta/f$ from the fiducial full-sky (no ZoA) case. The top (bottom) panels show the ratio for the $K < 12.5$ ($K < 11.5)$ flux-limited case. In all cases, we measure $\beta$ in radial bins of 25 \h Mpc and absolute Galactic latitude bins of $5^\circ$. We omit bins where any mock has less than 10 galaxies from this diagram.  
    The results for the deeper (top row) and shallower (bottom row) surveys are both shown here, with the dotted line representing the galactic cut applied to each.  The measurements performed below this line are calculated using velocity tracers that would have been observed in the absence of the galactic plane. For clarity, we note the printed values overlaid in each bin are simply the numerical value of the bin's measurement, which in turn corresponds to the colour bar.
    }
    \label{fig:betagrid}
\end{figure*}

\begin{figure*}
    \centering
    \includegraphics[width=\textwidth]{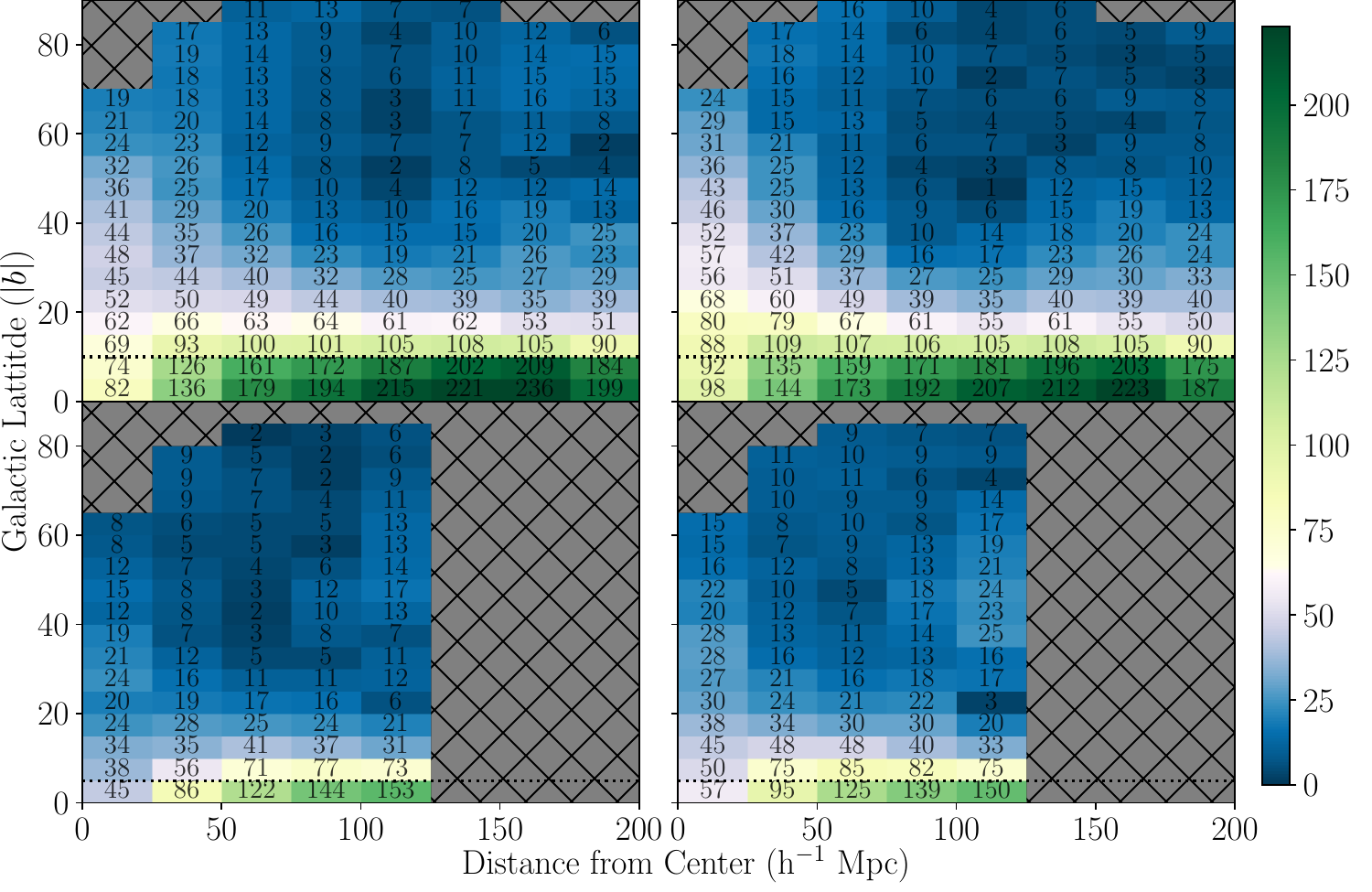}
    \caption{The binning and layout have the same meaning as in  Fig. \ref{fig:betagrid}, but here we show the additional velocity scatter that is generated due to treatment of the ZoA. Thus the total scatter, $\sigma\sbr{v}$, as a function of distance and Galactic latitude can be obtained by adding the values here in quadrature with the values from Fig.\ref{fig:12.5v11.5}. 
    } 
    \label{fig:sigvgrid}
\end{figure*}

Truly all-sky galaxy surveys are difficult to acquire due to the ZoA. In this region, dust extinction due to the dust in the disk of the Milky Way, as well as the high density of stars in this region, diminish our ability not only to identify galaxies but to accurately measure their properties.
Therefore, at these low Galactic latitudes, our ability to generate an accurate map of the large-scale structure impedes our ability to predict the peculiar velocities of nearby galaxies and thus introduces an added level of difficulty when measuring cosmological parameters. 

Recall that in 2M++ there are two galactic latitude cuts applied to the survey data, corresponding to the surveys used in this survey compilation. As the location of the Galactic plane is arbitrary in our mocks, to each mock we consider three different galactic plane orientations each one orthogonal to the other.  In this section, we explore the effects of a simple Galactic latitude cut.  We consider two cases ($|b| < 5$\degree\ and $|b| < 10$\degree) corresponding to the two parts of 2M++. The results for the mock 2M++ catalogues can be found in the following section.

\subsection{Treating the ZoA}\label{sec:ZoATreatment}

There are two simple methods to compensate for the ZoA when generating velocity fields: 1) filling the region with a uniform value, or 2) mimicking the missing structure with observed galaxies outside of the ZoA.  To emulate the former, we simply fill the ZoA region with the calculated mean density of the local Universe; this ensures that the value for this region is equivalent to the regions outside of our survey volume. In the latter case, in order to generate missing structure within the ZoA, we adopt a technique developed by \cite{lynden-bell_cosmological_1989}: and ``clone'' the galaxies that are immediately located above and below the ZoA. To generate a clone galaxy at a latitude of $b_c$ it is simply a matter of shifting the original latitude, $b$, using:
\begin{equation}
b\sbr{c}  = \arcsin[\sin(b_\text{ZoA})-\sin(b)]
\end{equation}

In order to fully understand the effects of the ZoA on the full 2M++ catalogue, we consider two galactic latitudes to define the ZoA. In the full 2M++ catalogue, the SDSS and 6dFGRS surveys ($K_{2M++} \leq 12.5)$ do not cover the region where $|b| < 10^\circ$, whereas for the shallower 2MRS ($K_{2M++} \leq 11.5$) this region is $|b| < 5^\circ$. As a result while 2M++ provides nearly full sky coverage, the survey conditions vary across the sky. In this section, however, we first consider near full-sky surveys with only simple latitude cuts. The realistic 2M++ scenario will be discussed in Section \ref{sec:ZoA2MPP}.

\subsection{Effects the ZoA}\label{sec:ZoA_results}

To demonstrate how our predictions are influenced by these Galactic latitude cuts, Fig.\ \ref{fig:betagrid} shows how the measurements of $\beta/f$ change as a function of both galactic latitude and distance from the centre of our mock Universes, for the flux-limited luminosity weighted samples.

We find that, on average, the cloning treatment tends to decrease the recovered value of $\beta/f$ near the Galactic plane, whereas the uniform treatment increases our recovered values. As the ratio is with respect to the full-sky case, the areas that are fully within the ZoA are strongly biased, as expected, but outside the ZoA these biases reduce to less than 10\% within $5-10 ^\circ$, depending on the size of the ZoA. Recovered values are not drastically impacted in regions where the Galactic latitude is greater than twice the applied cut's latitude. 

We find that the uniform treatment tends to cause significantly greater bias in the recovered measurements at nearby distances than the cloning treatment. 
For predictions in the nearby Universe (performed at distances within $50$ \h Mpc), the uniform treatment is more biased by  5-15\%, although in the opposite sense to the cloning case. Neither treatment is able to fully mitigate the increased inaccuracy in the velocity predictions due to the missing structures.  One might naively assume that introducing structure with the cloning treatment would improve recovered values, however there is no guarantee that the cloned structure perfectly matches the ``missing'' structure that is hidden by the Galactic plane. This leads to a bias in the recovered $\beta$ and an increase in the error in the predicted peculiar velocities, $\sigma\sbr{v}$.  We find no significant advantage to using one method over the other for predictions made \emph{within} the ZoA: in both cases, the total scatter can reach around 250 km s$^{-1}$, approaching the cosmic r.m.s.\ velocity. 

Both treatments for the ZoA result in comparable amounts of scatter being added as a function of latitude. Additionally, beyond twice the Galactic latitude cut, the amount of scatter added is small compared to the overall scatter in the predicted peculiar velocities.

This can also be seen in the velocity scatter shown in Fig.\ \ref{fig:sigvgrid}. This diagram has the same binning as in Fig.\ \ref{fig:betagrid}, and shows the additional scatter in velocity that is introduced solely as a result of the ZoA treatment used. It is clear that the extra scatter in predicted peculiar velocities as a result of the ZoA is comparable for both treatment methods. As expected, the scatter increases in the same regions where the predictions are most heavily impacted by the ZoA, as was discussed previously.  We note that the additional scatter introduced is still less than the predicted variations from linear theory alone. Even in the regions located directly above the ZoA, we expect the total scatter in our predictions to be only $\sim 30$ km s$^{-1}$ larger than for those that are far from the ZoA. For the regions within the ZoA, i.e., $b < b\sbr{ZoA}$, as discussed previously, the scatter is on average higher for cloning than the uniform treatment by $\sim 15$ km s$^{-1}$.

\section{Analysing the fully-realised 2M++ mocks}
\label{sec:combined}

There are two main differences between the analysis of the full 2M++ in C15 and what we have discussed so far: the treatment of the ZoA, which is discussed in Section \ref{sec:ZoA2MPP}, and the way in which density fluctuations are normalised, in Section \ref{sec:galaxybiasnorm} below.

\subsection{Treatment of the ZoA in 2M++}
\label{sec:ZoA2MPP}

In the previous sections, our measurements for each of the $K$-band limited cases had a constant Galactic latitude cut for the ZoA and single maximum distance. The full 2M++ case is more complicated: the ZoA is defined as $|b| < b\sbr{ZoA}$ where
\begin{equation}
b\sbr{ZoA}  =
\begin{cases}
  5^\circ & \text{if } 30^\circ < l < 330^\circ \\
 10^\circ & \text{if } l < 30^\circ  \text{~or~} l > 330^\circ \,,
\end{cases}
\end{equation}
with the additional complication that C15 only performs the cloning procedure out to a distance of 125 \h Mpc for galactic longitudes $30^\circ < l < 330^\circ$. At the longitudes close to the Galactic centre and in the distance range $125 \h \textrm{Mpc} < r < 200 \h \textrm{Mpc}$, the uniform ZoA procedure is used.

One of the goals in this work is to provide a more comprehensive uncertainty estimate of galaxies peculiar velocities as a function of their galactic coordinates and depth rather the uniform 150 km s$^{-1}$ proposed in C15. For this section, we will continue to use the same convention as in C15, to understand potential uncertainties that need to be accounted for in the 2M++ analysis. 
We find that, as expected, the recovered value of \bstar\ calculated using the convention described above 
produces a value of $1.03 \pm 0.05$, for a hypothetical peculiar velocity sample with the same redshift distribution as the 2M++ galaxies themselves. This value is in between that measured for the globally applied ZoA of $10^\circ$ for the 200 $h^{-1}$ Mpc mocks, as described in \ref{sec:ZoA_results}, which found values of \bstar\ of $0.97 \pm 0.04 $ for the cloned and $1.05 \pm 0.04$ for the uniform treatments. It is unsurprising that for this case the global 2M++ measurement of \bstar\ tends to align more with the latter case due to the geometry of 2M++ discussed previously. This remains true for the number-weighted 2M++ measurement of  $0.97 \pm 0.06$, which has a recovered cloned-ZoA and uniform-ZoA values of $0.93 \pm 0.05$ and $1.0 \pm 0.04$, respectively.
In either case the recovered value is comparable to the full-sky measurement for which $\bstar_{L_K} = 1.03 \pm 0.03$ and $\bstar_{N} =  0.99 \pm 0.05$

\subsection{Normalising the density field and the resulting fits of $\beta$ and $f\sig$}
\label{sec:galaxybiasnorm}

\begin{figure}
    \centering
    \includegraphics[width=.5\textwidth]{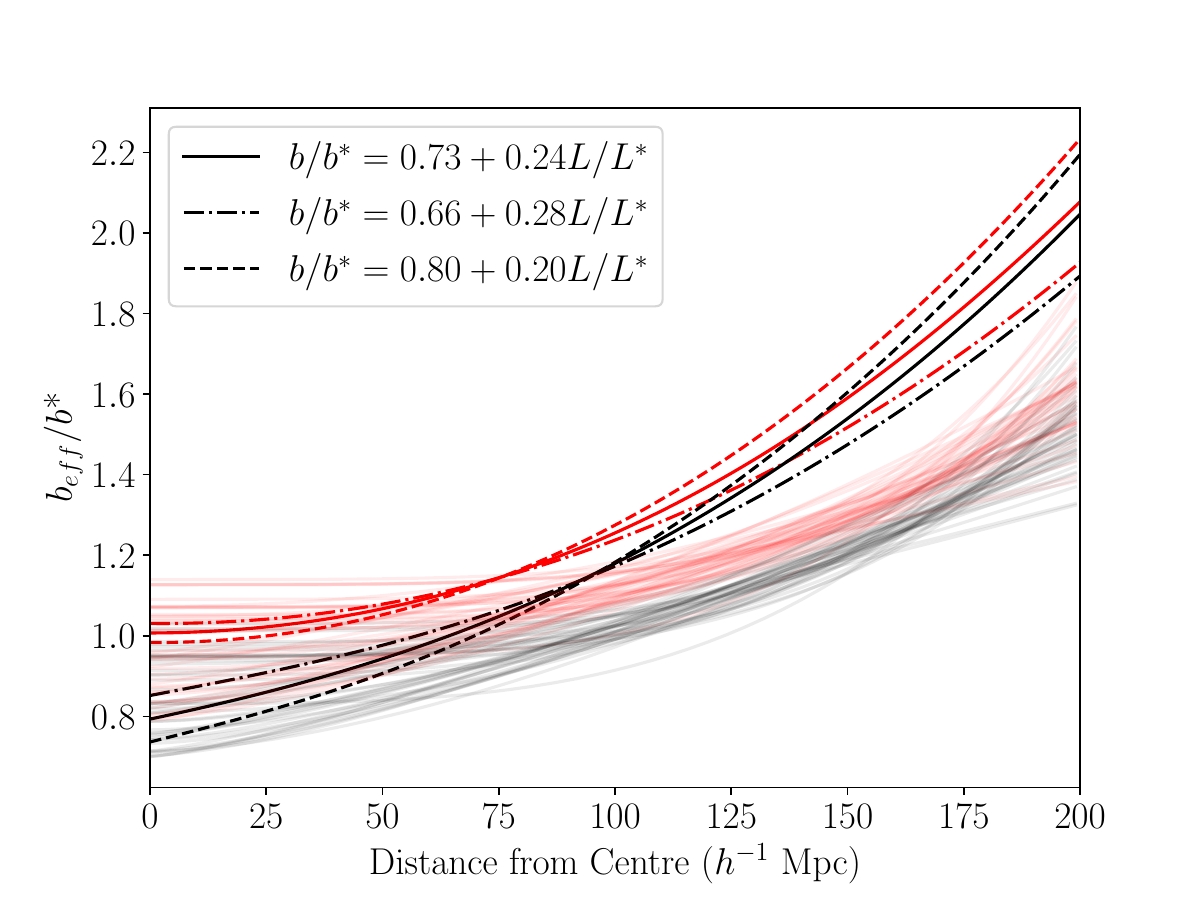}
    \caption{Number-weighted (black) and luminosity-weighted (red) effective bias as a function of distance for the most significant $1\sigma$ deviation of parameters from the scaling relation $b/b_* =(0.73 \pm 0.07) + (0.24 \pm 0.04) L/L_{*}$ from C15. 
    The thin transparent lines are the parametric fits F(r) 
    of $\sig$ as a function of distance from the centre of the sphere for each mock Universe.}
    \label{fig:Recreation_Carrick_fig3}
\end{figure}

For 2M++-like surveys which are magnitude limited, the mean luminosity of the galaxies observed increases with depth. This means that on average galaxies at higher redshifts are more biased than those which are nearby, as galaxy bias increases with luminosity. In C15, this was performed by re-scaling the density field to the same effective bias using the bias model of \cite{westover_galaxy_2007}, where bias changes with luminosity. That work found that $b(L)/b^* = (0.73 \pm 0.07) + (0.24 \pm 0.04)L/L^*$, where $b^*$ is the bias of an $L^*$ galaxy, and was determined using the correlation function of 2MASS. Since different distance ranges sample different luminosities, there is an effective bias at each distance which can calculated from the above. Fig.\ \ref{fig:Recreation_Carrick_fig3} shows the  functional form of $b\sbr{eff}(r)/b^*$ from \cite{westover_galaxy_2007} as calculated in C15. This was then used to normalise $\delta\sbr{g}$ so that the $\sigma\sbr{8,g}$ at all radii corresponds to that of an $L_*$ galaxy.

\begin{figure}
    \centering
    \includegraphics[width=.5\textwidth]{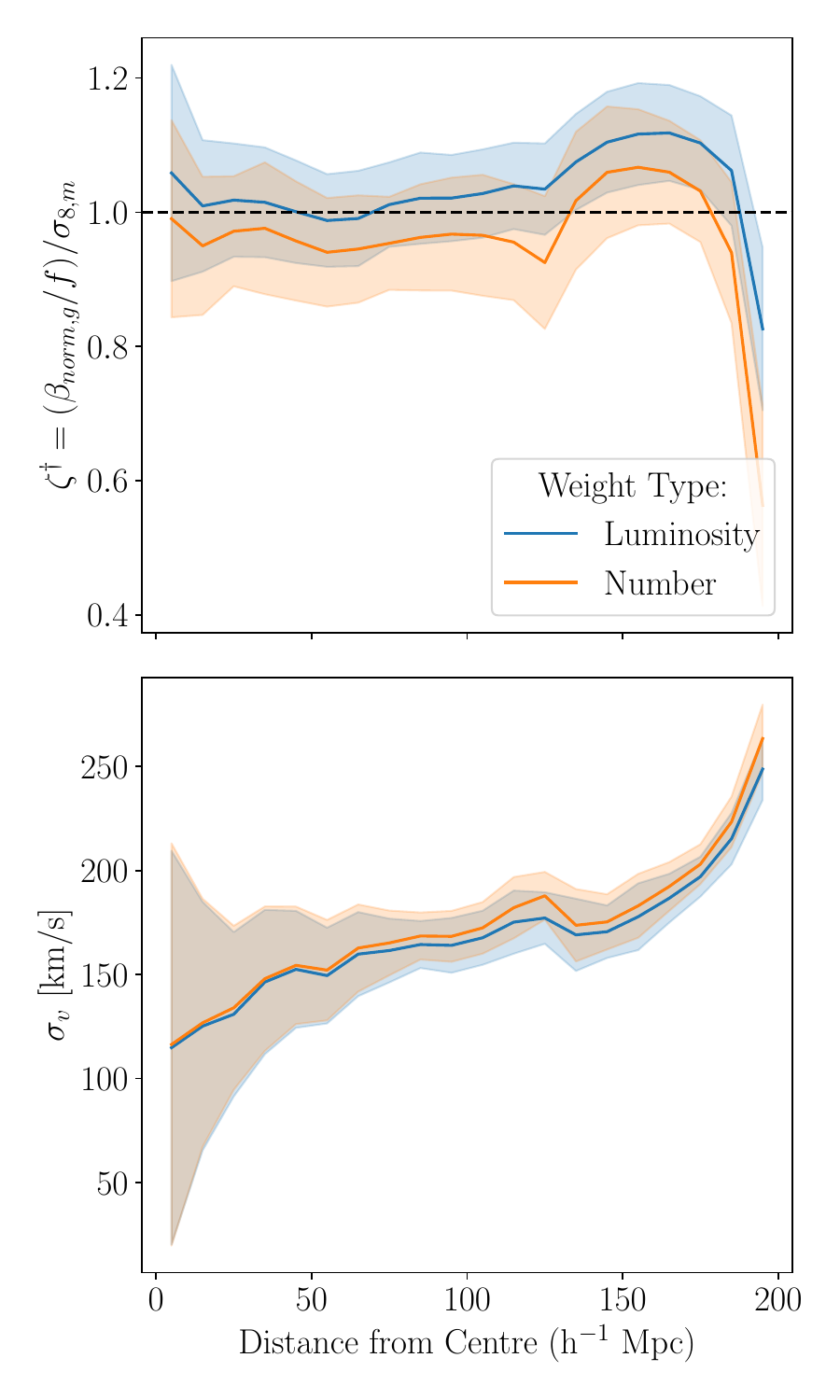}
    \caption{The bias in the normalised fits, $\zeta^{\dagger} = \beta\sbr{norm}/f/\sigma\sbr{8,m}$,  for the mock 2M++ realisations. The density field is corrected first by the individual galaxy weights, then normalised by the fitted \sig\ curves shown in Fig.\  \ref{fig:Recreation_Carrick_fig3}.}
    \label{fig:bstar}
\end{figure}

In this section, however, we use a slightly different method to normalise the density field. We directly re-scale the density field as follows. For each mock, we first calculate $\sig$ in equal volume shells as a function of depth, and fit a parametric form to it, denoted $F(r)$, these are shown as the lighter bands in Fig.\ \ref{fig:Recreation_Carrick_fig3}. We divide the density field with this normalisation: $\delta\sbr{norm}(r) = \delta\sbr{g}(r)/F(r)$. The regressions performed using $\delta\sbr{norm}$ thus gives a value of $\beta\sbr{norm}/f$ and, in analogy to $\beta^*$, we define the bias in this method as
\begin{equation}
    \zeta^\dagger \equiv \frac{\beta\sbr{norm}}{f \sigma\sbr{8,m}}\,.
\end{equation}
 
This normalisation procedure was applied to the 2M++ mocks. A summary of the global measurements produced using this method, again excluding survey edges,  can be found in Table \ref{table:Tab2}.
\begin{figure*}
    \centering
    \includegraphics[width=\textwidth]{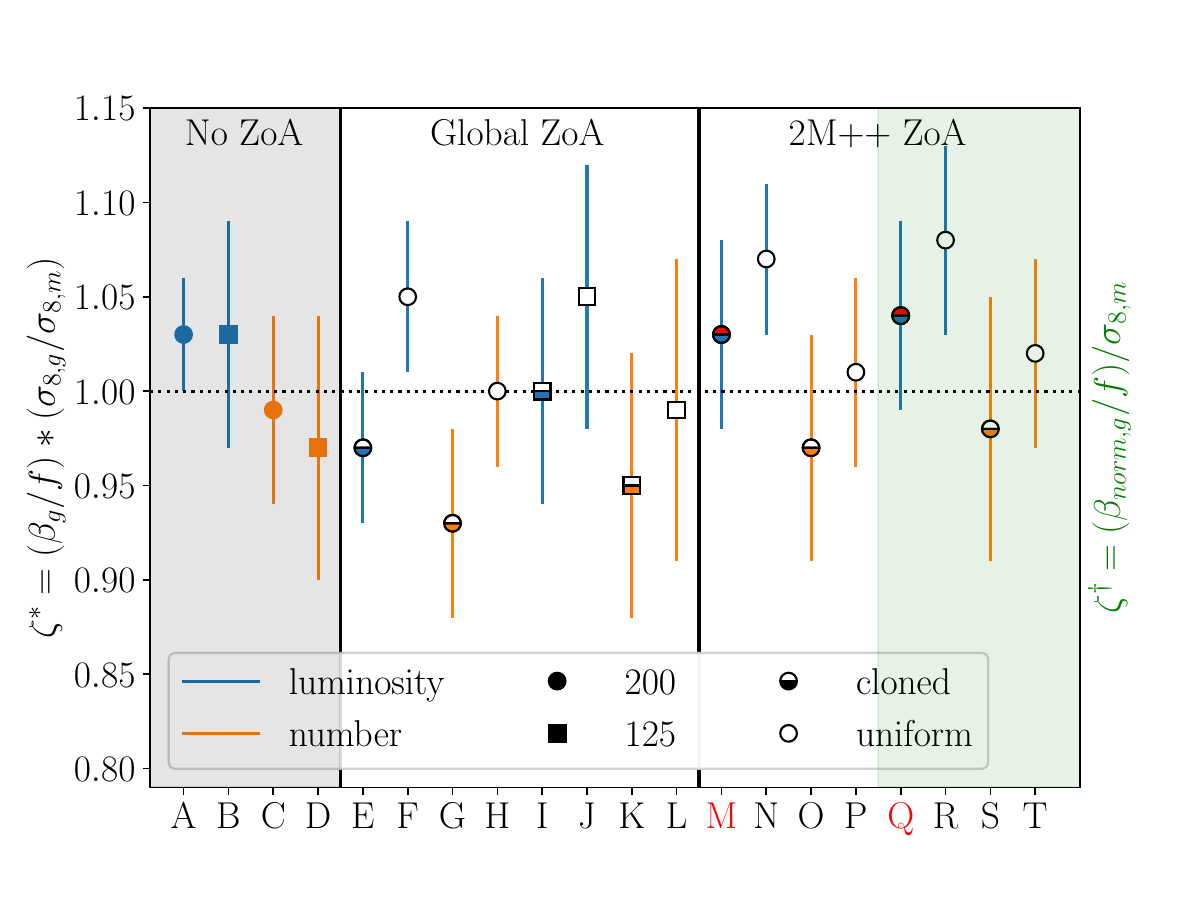}
    \caption{Visual summary of the more important values presented in Tables \ref{table:Tab1} and \ref{table:Tab2}. The letters A-P correspond to the values of \bstar\ in \ref{table:Tab1}. A-D represent the flux and volume limited full-sky measurements, while E-L are the values calculated assuming a single global latitude cut corresponding to the two survey depths used in 2M++. Values for M-P are calculated using the fully realised 2M++ mocks. Finally Q-T are the $\zeta^{\dagger}$ values described in \ref{table:Tab2} of the 2M++ mocks. The blue and orange lines represent the density weighting used in calculation, while the symbol shapes are the maximum survey distance cut allowed, and the fill choice of the markers represent the method used to treat the ZoA; full: no ZoA applied, half filled: cloning the structure above and below the ZoA, empty: uniform value applied to the ZoA. The scatter in the values is the cosmic variance between the different realisations.}   
    \label{fig:summary}
\end{figure*}

As can be seen, the bias-normalisation correction in C15 is greater in the outer regions of the survey than our fitted measurements of $\sig$ indicate. This is likely due to the fact that the SAMs applied to MDPL2 do not quite reproduce the same level of clustering as the real universe, as has also been found in previous works \citep{knebe_multidark-galaxies_2018, lilow_constrained_2021}. We also find that SAGE shows significantly less clustering than SAG. This indicates that the mocks underestimate the $\sigma_g$ of the real Universe, in spite of having a higher $f\sig$ than the real Universe. Should the simulations better reflect the clustering of the Universe, we should get the same correction as in \cite{westover_galaxy_2007}. While the Westover correction function would significantly over-correct our mocks generated by SAG and SAGE, it not obvious that this would over-correct the real Universe. In Fig.\ \ref{fig:bstar}, we show our measurement of $\zeta^\dagger$ for the realistic luminosity- and number-weighted 2M++ mocks. Overall our corrections, applied to both the luminosity and number weighted fields, result in a $\zeta^\dagger$ which is in good agreement with unity across radial shell measurements. In particular, we see that both SAG and SAGE have very similar values of  $\zeta^\dagger$, in spite of the differences in their clustering  ($\lesssim 1\%$, see Table \ref{table:Tab2}). This suggest that, even though the real Universe is more clustered than either of these SAMs, our procedure is insensitive to the exact level of galaxy clustering.   
We note however,  that while the overall external bulk flow and global $\zeta^\dagger$ calculated excludes the outer survey edge, the same is not true here. Hence as expected, there are two distances where the recovered $\zeta^\dagger$ is affected by the survey edges, corresponding to 125 and 200 \h Mpc.

Fig.\ \ref{fig:summary} presents a visual summary of the key calculations performed in this work. The red points represent our measurements on the fully realised luminosity-weighted 2M++ mocks. For these we  find that our treatments are   comparable to the full-sky case (A). The realistic cloning treatment (M) for the ZoA is a marginally better treatment than the fully uniform case (N), returning a more comparable value to the our predictions for the  full-sky  case (A). However both treatments are in agreement with the fiducial value. Contrarily, on average for the number weighted case  the uniform treatment gives a more accurate reconstruction to the fiducial, though again both are well within the uncertainty margins.  We find that for values measured using the partial 2M++ cloning treatment, the values of \bstar(M) and  $\zeta^\dagger$ (Q) are almost equivalent, however the latter tends to produce a slightly higher measurement.  The same is true for the fully uniform treatment, shown by points (N) and (R). 

\begin{figure*}
    \centering
    \includegraphics[width=\textwidth]{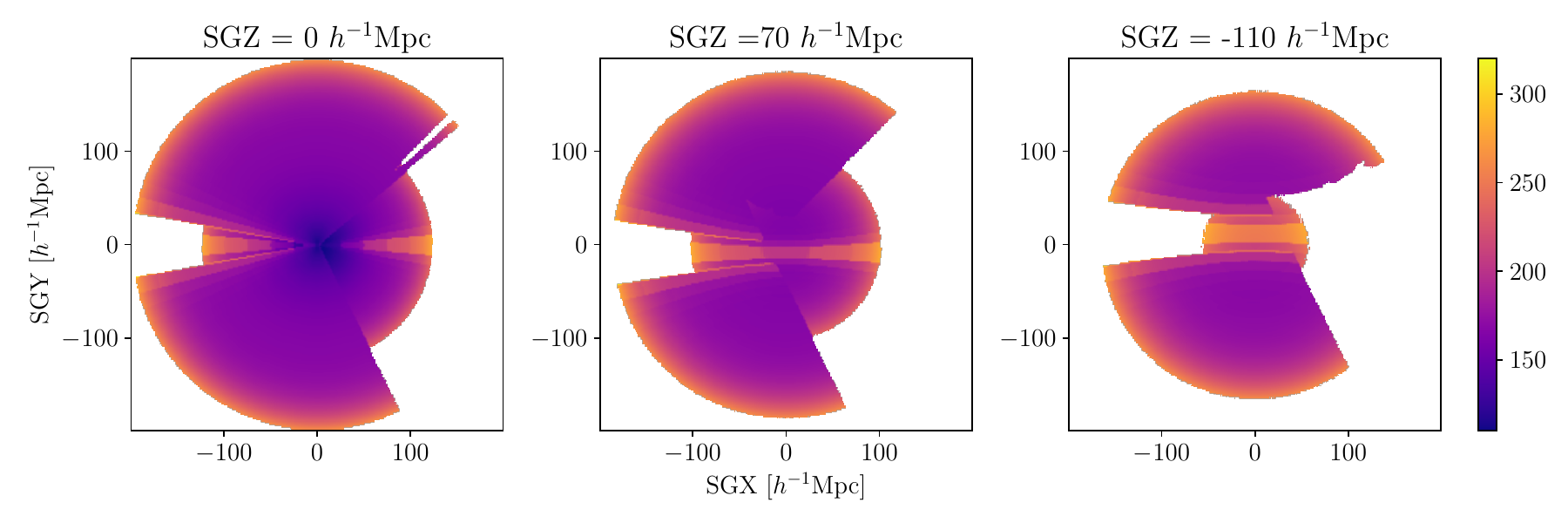}
    \caption{The total error, $\sigma\sbr{v}$, in the predicted peculiar velocities from the 2M++ luminosity-weighted galaxy density contrast field,  using the cloning treatment of C15, for various slices in the Supergalactic Plane ($SGZ = 0$), and parallel to it. 
    The latter are a select sample of those depicted in C15. The 2M++ density field is shallower at positive SGX as this region is covered by the 2MRS survey, whereas the remainder of the sky is covered by the deeper 6dFGRS and SDSS.}
    \label{fig:pverr}
\end{figure*}

Fig.\ \ref{fig:pverr} demonstrates how the scatter in our predictions changes as a function of distance from the origin and galactic latitude. Shown are slices in the Supergalactic coordinate frame at SGZ =0, +70 and -100 \h Mpc for the realistic 2M++ treatment of luminosity weighted field. These slices demonstrate how predictions are impacted by both the shallower 2MRS regions (at positive SGX), and the deeper regions covered by SDSS and 6dFGRS. We add the average velocity scatter and the excess scatter due to the ZoA treatment in quadrature, separately for these regions as predictions are most heavily influenced by their depth (see Fig. \ref{fig:12.5v11.5}) and galactic latitude above the ZoA (as shown in Fig. \ref{fig:betagrid}). As expected the scatter in predictions are highest in regions obscured by the galactic plane where the no structure data is known at $\sim 240$ km s$^{-1}$. In the regions above and below the galactic plane, the scatter ranges between 120-260 km s$^{-1}$ with galaxies on average having a scatter of $\sim 170$ km s$^{-1}$. This is larger than the value adopted by C15 of 150 km s$^{-1}$ but not as extreme as the 250 km s$^{-1}$ used in recent supernova analyses \cite[such as in][]{riess_new_2018, scolnic_complete_2018, kenworthy_measurements_2022, riess_comprehensive_2022}. Again, we note that these estimates are only valid for central galaxies, as linear perturbation theory does not accurately predict the additional virialized velocities of satellite galaxies with respect to the group/cluster's centre of mass.

\section{Impacts on Previous Results}\label{sec:impacts}

\subsection{Corrected measurements of $\beta$ and $f\sig$}

Having carefully calibrated the bias $\zeta^{\dagger}$ as a function of distance (the solid lines in Fig.\ \ref{fig:bstar}), and its cosmic variance (the shaded region in Fig.\ \ref{fig:bstar}), we can use this to re-calibrate previous results that have used 2M++ to estimate 
\begin{equation}
f\sig = \beta\sbr{norm}/\zeta^{\dagger} \,,
\end{equation}
and their cosmic-variance driven uncertainties.

We focus on two of the most recent of such studies, namely \cite{boruah_cosmic_2020} and \cite{said_joint_2020}. We divide the peculiar velocity data used \cite{boruah_cosmic_2020}  into four subsamples: one of Type Ia supernovae (labelled ``A2'' in that paper), and three Tully-Fisher-based subsamples (SFI Field galaxies, SFI group galaxies and 2MTF galaxies).  Likewise, we divide that in \cite{said_joint_2020} into two Fundamental Plane (FP) samples (labelled 6dF and SDSS). The mean corrections for each of these six subsamples are small: when weighting these samples in the same way as is done when fitting the peculiar velocity data, we find weighted-average values of $\zeta^{\dagger} = 1.01, 1.01, 1.01, 1.01, 1.04, 1.04$ for the six subsamples respectively. After making these corrections, we can re-derive new, bias-corrected values of $\beta$ for each sample. 

To combine these to yield $f\sig$ is more difficult, because in general they are not completely independent. The peculiar velocity measurement errors on the individual $\beta$ values are independent, but when placed into the cosmic context and converted to values of $f \sigma_8$ it becomes more complicated. As discussed in HH21, the uncertainty in the value of $f\sig$ is driven primarily by the finite volume of 2M++ and cosmic variance. In particular, the density fluctuations are normalised within the 2M++ volume (effectively a sphere of 150 Mpc/h). If, for example, the 2M++ volume happened to be under-dense compared to the true mean density, then the local value of $\bar{\rho}$ would be too low, resulting in derived (locally-normalised) values of $\delta$ that are too high, leading to a fitted $\beta$ that is too low for our ``local'' sphere. Moreover, because of sample variance, the local \sig\ may also differ from the global one. The 2M++ mocks are embedded in a 1 \h Gpc simulation box, so the variations in local mean density and \sig\ will be captured in the mocks, as will the effects of shot noise and ZoA cuts. Table \ref{table:Tab2} shows that for both SAMs, the standard deviation from mock-to-mock is 5\%. In C15, this was estimated at 4\% from the scatter in different shells, in reasonable agreement with the mock-to-mock variance calculater here. This cosmic variance is the dominant source of error in $f\sig$. But in addition, because the individual peculiar velocities sample different parts of the local sphere there is additional sample variance. For example, the \cite{boruah_cosmic_2020} samples cover the sky outside the ZoA but are relatively shallow, with characteristic depths ranging from 22 to 40 \h Mpc. On the other hand the FP samples in \cite{said_joint_2020} are deeper (typically 100 \h Mpc) but cover different parts of the sky (the southern hemisphere in the case of 6dF, and a smaller 1 steradian cap in the North for SDSS). When we compare these we find variations in $\beta$ that are larger than just the measurement errors. We attribute this to sample variance effects.

We consider three different methods for combining these measurements. The first method is to include an additional error in quadrature so that the $\chi^2$ of the weighted mean $\beta$ of the six samples is equal to the number of degrees of freedom (5). This method yields $\beta = 0.390\pm0.024$. A separate method is to ``jackknife'' over the six surveys,  which yields $\beta = 0.402\pm0.022$.  Finally, we estimate the covariance of the individual $\beta$ measurements using the mocks. This yields $0.402\pm0.023$. These three methods are in good agreement with each other in terms of their mean values and uncertainties. We adopt a best estimate of $f\sig = 0.398\pm0.025$.

The above result is the non-linear value of $\sig$. To compare with cosmological models, it is better to have the linear value of \sig. In previous work, we used the prescription of \cite{juszkiewicz_weakly_2010}, which is based on 2nd-order perturbation theory, to correct from the non-linear \sig\ to the appropriate linear value. Here instead we use \texttt{halofit} \citep{smith_stable_2003, takahashi_revising_2012} which better matches the measured particle $\sig$ in MDPL2 (see Section \ref{sec:sim_data}). With \texttt{halofit}, we obtain a larger correction than in our previous work, and hence a lower $f\sigma_8^{\textrm{lin}} = 0.362\pm0.023$.

Our value of $f\sigma_8^{\textrm{lin}}$ is in good agreement with that of \cite{lilow_constrained_2021}, who found $f\sigma_8^{\textrm{lin}} = 0.367\pm0.060$ using the density field from the slightly shallower 2MRS survey \citep{huchra_2mass_2012}, the Cosmic-Flows 4 peculiar velocity compilation \citep{tully_cosmicflows-4_2023} and which uses a different method for comparing the fields.

Assuming $\Lambda$CDM, our result translates to $S_8 \equiv (\Omega\sbr{m}/0.3)^{0.5} \sigma_8^{\textrm{lin}} = 0.702\pm0.044$. which deviates from the \cite{planck_collaboration_planck_2020-1} result at the $2.8 \sigma$ significance level, but is in agreement with the lower measurements from weak gravitational lensing of galaxies \citep{heymans_cfhtlens_2013, asgari_kids-1000_2021, heymans_kids-1000_2021, abbott_dark_2022, dalal_hyper_2023}. Peculiar velocities are complimentary to the weak lensing of galaxies, and have the advantage that the predicted velocities, which are generated by a density field already smoothed by a 4 \h Mpc Gaussian filter, are exceedingly insensitive to non-linearities. Indeed, the predicted root-mean-square peculiar velocity drops by only 1.2\% if a linear power spectrum is used in place of a non-linear one. In the context of weak lensing, for example, \cite{amon_nonlinear_2022} have suggested that lowering the non-linear contribution to the matter power spectrum by a factor $A_{\rm mod} = 0.69\pm0.04$ is a possible way to reconcile the weak gravitational lensing results with those from \cite{planck_collaboration_planck_2020-1}, which probes larger, linear scales. Because peculiar velocities are sensitive primarily to large scales, and the predictions are smoothed on scales of 4 \h Mpc, such a correction would only reduce the typical peculiar velocity by less than 1\% and so would have no significant effect on our calculated values of $\beta$. In summary, the contribution from non-linear scales to peculiar velocities is negligible. This suggests that a scale dependent correction to the matter power spectrum is not the solution to the ``$S_8$ tension.''

 Peculiar velocities probe the matter power spectrum at redshifts very close to zero ($z \sim 0.03$). Current galaxy weak lensing surveys are sensitive to low redshifts ($z \sim 0.4$), whereas CMB lensing is most sensitive to redshifts greater than one, and the primordial CMB anisotropies are sensitive to structures at $z \sim 1100$. Peculiar velocities and weak gravitational lensing predict lower values of $S_8$, whereas CMB lensing \citep{planck_collaboration_planck_2020-7, pan_measurement_2023, madhavacheril_atacama_2024}, and the primordial anisotropies \citep{planck_collaboration_planck_2020-1} yield higher values. This has led to the suggestion that the growth of matter density fluctuations has a different redshift dependency than that expected in the $\Lambda$CDM model. Recall that the amplitude of fluctuations, $D$, depends on the scale factor, $a$, as $d \ln D/d\ln a = \Omega_{\rm m}^{\gamma}$, with $\gamma = 0.55$  in $\Lambda$CDM. If we assume that the growth index $\gamma$ differs from $\Lambda$CDM, but that \cite{planck_collaboration_planck_2020-1} measurements are correct for high redshifts, and our values are accurate at low redshifts, then our fits require $\gamma = 0.699\pm0.057$, a $2.6 \sigma$ deviation from the $\Lambda$CDM expectation of $\gamma = 0.55$. This is consistent with other studies that have used direct peculiar velocity measurements, in whole or in part, to constrain $\gamma$ \citep{hudson_growth_2012, johnson_searching_2016, said_joint_2020, nguyen_evidence_2023}, although the peculiar velocity datasets have some overlap and hence are not completely independent.

\subsection{External Bulk Flows}\label{sec:bulkflow}

It is expected that, as peculiar velocity surveys extend to larger and larger scales, the bulk flow of the surveyed volume, with respect to the rest frame defined by the CMB, will approach zero. The degree to which it converges as a function of volume, however, is a measure of the power spectrum on large scales. Over the last 15 years, there have been a number of studies of the bulk flow: some of these have found bulk flows higher than expected in $\Lambda$CDM, while others have not found any  statistically significant discrepancy. These studies simply take a weighted average of the measured peculiar velocities.

The quantity measured in this work, however, is the residual or external bulk flow, $\boldsymbol{V}\sbr{ext}$ (see equation (\ref{eqn:v_r_density})). In principle, this can also be used to assess the power spectrum on scales larger than the survey limit, but the interpretation is more complicated than for a simple bulk flow. If the density field had had a simple geometry, such as a sphere, and no noise, it would have been straightforward to calculate the cosmological expectation the external bulk flow. For example, for a 200 \h Mpc sphere, and for a $\Lambda$CDM model scaled to $\sig = 0.8$, \cite{hudson_streaming_2001} found $\sim 40$ km s$^{-1}$ in each Cartesian component.  In the presence of an unusual geometry and noise, however, we must use the mocks to assess a range of external bulk flows that are expected.

\begin{figure}
    \centering
    \includegraphics[width=\columnwidth]{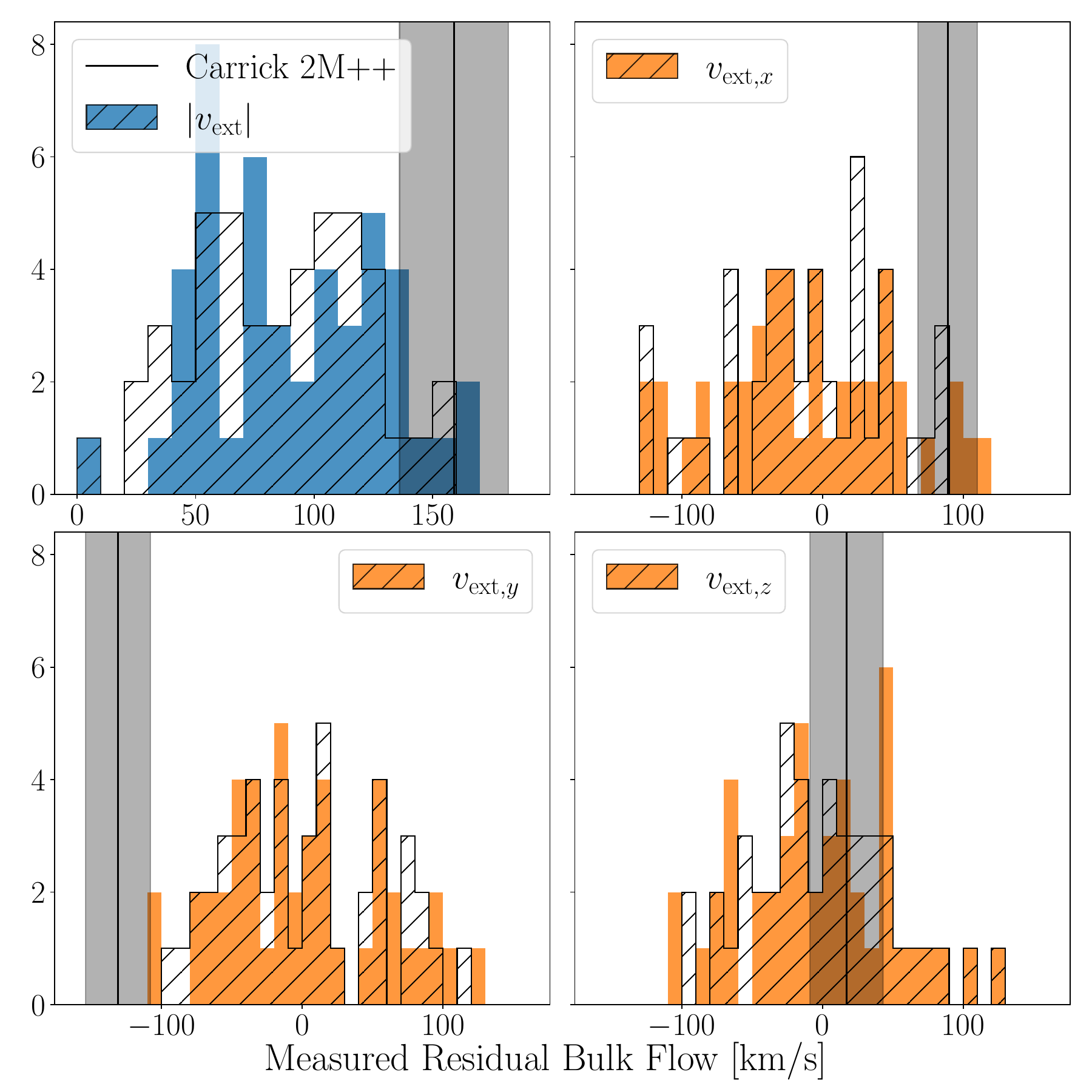}
    \caption{Top left: Shown in blue is a histogram of the magnitude of the measured $\zeta^\dagger$ residual bulk flow of the for the mock catalogues with the full 2M++ conditions imposed, the hatched histogram overlaid shows the same calculation performed instead by \bstar.  The black vertical bar with grey error bands serves as a comparison to the measured 2M++ value from \protect{C15}. The orange histograms show the individual Cartesian components of the external flow. The $Z$-component demonstrates slightly less variance because it is less affected by the cloning of the ZoA.}
    \label{fig:bulkflow}
\end{figure}

In Fig. \ref{fig:bulkflow}, we show the measured external bulk flow from our mock catalogues, again excluding the outer 20 \h Mpc edges where predictions are heavily influenced by the lack of structure outside of survey limits. The black line shows the external bulk flow calculated in C15. The blue histogram shows the absolute magnitude of the external bulk flow, calculated from our $\zeta^\dagger$ values, while the orange histograms show the measured Cartesian components. For comparison the hatched values show the measurements of \bstar. The standard deviations of the Galactic Cartesian components external velocity flows using the $\zeta^{\dagger}$ method are respectively 60, 56 and 54 for the $X$, $Y$ and $Z$ components. 
As expected, since the missing structure is in the Galactic plane, the standard deviation of the $Z$ component is smaller than that of the $X$ and $Y$ components which are of the same order due to symmetry.

It is interesting to check whether the observed external bulk flow found by C15 is consistent with the expectation from $\Lambda$CDM. In $\Lambda$CDM, we expect the ensemble of such spheres to have a mean of zero, but with a standard deviation that we can calculate from the mocks as discussed above.   Combining the individual Cartesian components' observational error from C15 (their Table 3) and adding the observational uncertainties in quadrature with the cosmic standard deviation, we obtain a $\chi^2 = 6.4$, which, when compared with the distribution with 3 degrees of freedom (for the 3 Cartesian components), yields a $p$-value of 0.094. This is below the threshold of what might be considered marginally significant.
Repeating this exercise for the external bulk flows 
found by \cite{boruah_cosmic_2020}, and for the 6dF and SDSS subsamples in \cite{said_joint_2020}, we find $p$-values of 0.06, 0.08 and 0.03, respectively. Of course, the external bulk flows from each of these samples are, to first order, responding to the same large-scale structures beyond the 2M++ volume and so are not independent, but are highly correlated.
Overall, there is therefore no conflict between the external bulk flows measured by these surveys and the expectations of $\Lambda$CDM. 

\section{Summary and Discussion}\label{sec:discussion}

Using the MDPL2 N-Body simulation and SAM stellar masses, we generated 15 independent \kband\ mock catalogues, from both SAG and SAGE. Then from each of these mocks, we generate catalogues with orthogonal rotations which mimic the selection and geometry of 2M++.
We constructed galaxy density fields of within 200 \h Mpc and, under the assumption of linear perturbation theory, predicted the peculiar velocities, which we then compared to the true known velocities from the simulation. The key results are as follows:

\begin{itemize}
\item 
We find that the predicted peculiar velocities for galaxies located within 20 \h Mpc of survey edges are less accurate (the velocity scatter in this region increases by $\sim 75$ km s$^{-1}$), impacting measurements of $\beta$ and reducing it by $20-30$\%. Away from the survey edges, however, the predictions work well.

\item 
Overall, we find that both the cloned and uniform treatments for the ZoA affect the recovered measurements of $\beta$. However the global cloning treatment performs marginally better. We find that original predicted values are effectively recovered to within 5\% for Galactic latitudes beyond twice the size of the applied ZoA.  

\item 
The results for \bstar and $\zeta^\dagger$ calculated using the same flux limited tracers for the velocity field, are summarised in Tables \ref{table:Tab1} and \ref{table:Tab2}. 
Measurements of \bstar and $\zeta^\dagger$ remain fairly constant across radial shell measurements, with the exception of the measurements made within $\leq 20 \h$ Mpc of the survey edges.

\item 
The systematic bias in $\beta$ for the ensemble of our 2M++-like mocks has been calibrated to a 1\% uncertainty. However, due to cosmic variance, the uncertainty for any single mock and for the real 2M++ is 5\%.

\item 
We find that the velocity scatter, $\sigma\sbr{v}$, of 150 km s$^{-1}$ adopted in C15 is slightly lower than the global measured scatter of $167 \pm 8$ km s$^{-1}$ for 2M++, but the predicted scatter is heavily influenced by both the distance to survey edge and proximity to the ZoA.

\item 
In our previous work (HH21) we predicted that cosmic variance would impact measurements of $f\sigma_8$ for a survey of 2M++'s size at the level of 5\%. Here we find consistent results. This can be mitigated with future deeper redshift surveys which probe larger volumes. The larger volumes have two advantages. First, their local mean density will have smaller fluctuations with respect to the global mean density. Second, the larger volumes will allow a better calibration of the local value of $\sigma\sbr{8,g}$.

\item 
The measured external bulk flow speeds found in C15 and \cite{boruah_cosmic_2020}, while higher than average, are still within the 95\% confidence range of what is expected due to cosmic variance in $\Lambda$CDM.  We note that the presence of the non-negligible external bulk flows are a result of the shallow depth of 2M++. 

\item Having calibrated the small bias of this method, we find $f\sigma_8^{\textrm{lin}} = 0.362\pm0.023$, or equivalently $S_8 = 0.702\pm0.044$, combining previous measurements based on 2M++ predicted peculiar velocities. This is lower than the predictions from Planck but in agreement with other peculiar velocity work and other techniques such as weak gravitational lensing.

\end{itemize}


\section*{Acknowledgements}


MJH acknowledges support from an NSERC Discovery grant.

The CosmoSim database used in this paper is a service by the Leibniz-Institute for Astrophysics Potsdam (AIP). The MultiDark database was developed in cooperation with the Spanish MultiDark Consolider Project CSD2009-00064. The authors gratefully acknowledge the Gauss Centre for Supercomputing e.V. (www.gauss-centre.eu) and the Partnership for Advanced Supercomputing in Europe (PRACE, www.prace-ri.eu) for funding the MultiDark simulation project by providing computing time on the GCS Supercomputer SuperMUC at Leibniz Supercomputing Centre (LRZ, www.lrz.de).

\section*{DATA AVAILABILITY STATEMENT}
The  data  underlying  this  article are publicly  available  from  the  COSMOSIM  database \url{https://www.cosmosim.org/}, with their respective publications cited in section \ref{sec:sim_data}.




\bibliographystyle{mnras}
\bibliography{PVs} 




\appendix
\section{\bstar and $\zeta^\dagger$ Values Measured in this work}

Here we present the full results for the measurements of \bstar (Table \ref{table:Tab1}) and  $\zeta^\dagger$  (Table \ref{table:Tab2}) for a wide range of different flux limits, weighting schemes and treatments of the ZoA.

\include{tables}


\bsp	
\label{lastpage}
\end{document}

%% file: tables.tex
\begin{table*}\caption{Measured values of \bstar\ and $\sigma_v$ for various tracers of our mock 2M++ catalogues. The row sections are categorised as follows: The first corresponds to the full-sky, the second to the globally applied ZoA cuts, and third the realistic 2M++ geometry. While the columns are:\\
1: The value corresponding to Fig \ref{fig:summary}. \\  
2: The maximum sphere radius applied. \\
3: The tracers ($t$) with which the underlying density field was generated, in all cases the velocity tracers are only central galaxies which made the flux limit cut.\\ 
4: Whether or not the density field tracers were limited to the flux limited cases, if the additional weighting for luminosity or number respectively was applied. \\ 
5: The region the ZoA was applied, in general 200 and 125 correspond to 10 and 5, 2M++ corresponds to the 2M++ realistic ZoA application. \\
6: The method used to treat the ZoA.\\
7: The measured value of $\bstar = (\beta\sbr{t}/f)(\sigma_8/\sigma\sbr{8,m})$ and the standard deviation from mock to mock for the SAG catalogues. \\
8: The velocity scatter $\sigma_v$ in our predictions for the SAG catalogues. \\
9: Same as 7 but for the SAGE catalogues \\
10: Same as 8 but for the SAGE catalogues}
\label{table:Tab1}
\begin{tabular}{llllllllll}
& max R &  & flux & Applied& ZoA  &  SAG  &  SAG  &  SAGE  &  SAGE   \\
& [$\h$ Mpc] & tracer type & limited &  ZoA & Treatment &  $\bstar$  &  $ \sigma_v$ &  $\bstar$ & $ \sigma_v$ \\ \hline
 & 200 & halo & False & -- & n/a & 1.07 $\pm$ 0.05 & 164.0 $\pm$ 6.4 & 1.09 $\pm$ 0.05 & 146.9 $\pm$ 4.8  \\
 & 200 & luminosity & False & -- & n/a & 1.04 $\pm$ 0.04 & 159.0 $\pm$ 5.9 & 1.04 $\pm$ 0.04 & 141.3 $\pm$ 4.2  \\
A & 200 & luminosity & True & -- & n/a & 1.03 $\pm$ 0.03 & 160.5 $\pm$ 6.2 & 1.04 $\pm$ 0.03 & 145.9 $\pm$ 4.1   \\
C & 200 & number & True & -- & n/a & 0.99 $\pm$ 0.05 & 163.9 $\pm$ 6.3 & 0.99 $\pm$ 0.06 & 147.5 $\pm$ 4.6   \\
 & 125 & halo & False & -- & n/a & 1.04 $\pm$ 0.06 & 162.4 $\pm$ 14.3 & 1.06 $\pm$ 0.07 & 146.4 $\pm$ 10.9   \\
 & 125 & luminosity & False & -- & n/a & 1.04 $\pm$ 0.05 & 157.7 $\pm$ 12.9 & 1.04 $\pm$ 0.05 & 141.5 $\pm$ 8.9   \\
B & 125 & luminosity & True & -- & n/a & 1.03 $\pm$ 0.06 & 159.7 $\pm$ 12.4 & 1.04 $\pm$ 0.05 & 145.2 $\pm$ 8.9  \\
D & 125 & number & True & -- & n/a & 0.97 $\pm$ 0.07 & 163.2 $\pm$ 11.9 & 0.99 $\pm$ 0.06 & 146.8 $\pm$ 8.3  \\
\hline
& 200 & halo & False & $10^\circ$ & cloned & 1.0 $\pm$ 0.05 & 170.5 $\pm$ 7.1 & 1.02 $\pm$ 0.05 & 154.0 $\pm$ 5.8  \\
 & 200 & luminosity & False & $10^\circ$ & cloned & 0.98 $\pm$ 0.04 & 166.0 $\pm$ 6.7 & 0.98 $\pm$ 0.04 & 148.9 $\pm$ 5.3  \\
E & 200 & luminosity & True & $10^\circ$ & cloned &  0.97 $\pm$ 0.04 & 167.7 $\pm$ 7.1 & 0.98 $\pm$ 0.04 & 153.4 $\pm$ 5.2 \\
G & 200 & number & True & $10^\circ$ & cloned &  0.93 $\pm$ 0.05 & 170.8 $\pm$ 7.2 & 0.93 $\pm$ 0.06 & 154.8 $\pm$ 5.8 \\
 & 200 & halo & False & $10^\circ$ & uniform & 1.08 $\pm$ 0.05 & 170.8 $\pm$ 7.4 & 1.1 $\pm$ 0.06 & 154.4 $\pm$ 6.0 \\
 & 200 & luminosity & False & $10^\circ$ & uniform & 1.06 $\pm$ 0.04 & 165.7 $\pm$ 6.6 & 1.06 $\pm$ 0.04 & 148.7 $\pm$ 5.0 \\
F & 200 & luminosity & True & $10^\circ$ & uniform & 1.05 $\pm$ 0.04 & 167.3 $\pm$ 6.9 & 1.06 $\pm$ 0.04 & 153.2 $\pm$ 4.9  \\
H & 200 & number & True & $10^\circ$ & uniform & 1.0 $\pm$ 0.04 & 171.5 $\pm$ 7.0 & 1.01 $\pm$ 0.05 & 155.7 $\pm$ 5.3 \\
 & 125 & halo & False & $5^\circ$ & cloned & 0.99 $\pm$ 0.07 & 165.9 $\pm$ 14.4 & 1.01 $\pm$ 0.07 & 150.0 $\pm$ 11.1 \\
 & 125 & luminosity & False & $5^\circ$ & cloned & 1.01 $\pm$ 0.05 & 160.5 $\pm$ 13.0 & 1.02 $\pm$ 0.05 & 144.3 $\pm$ 8.9  \\
I & 125 & luminosity & True & $5^\circ$ & cloned & 1.0 $\pm$ 0.06 & 162.6 $\pm$ 12.6 & 1.01 $\pm$ 0.05 & 148.2 $\pm$ 8.7 \\
K & 125 & number & True & $5^\circ$ & cloned & 0.95 $\pm$ 0.07 & 165.9 $\pm$ 12.0 & 0.97 $\pm$ 0.07 & 149.7 $\pm$ 8.1 \\
 & 125 & halo & False & $5^\circ$ & uniform & 1.05 $\pm$ 0.07 & 166.6 $\pm$ 14.8 & 1.07 $\pm$ 0.08 & 150.7 $\pm$ 11.3 \\
 & 125 & luminosity & False & $5^\circ$ & uniform & 1.07 $\pm$ 0.06 & 161.6 $\pm$ 13.3 & 1.07 $\pm$ 0.06 & 145.5 $\pm$ 9.2  \\
J & 125 & luminosity & True & $5^\circ$ & uniform & 1.05 $\pm$ 0.07 & 163.8 $\pm$ 13.2 & 1.07 $\pm$ 0.06 & 149.3 $\pm$ 9.3 \\
L & 125 & number & True & $5^\circ$ & uniform & 0.99 $\pm$ 0.08 & 168.6 $\pm$ 13.2 & 1.01 $\pm$ 0.07 & 152.6 $\pm$ 9.6  \\
\hline
 & 200 & halo & False & 2M++ & cloned & 1.05 $\pm$ 0.06 & 170.8 $\pm$ 8.1 & 1.06 $\pm$ 0.06 & 153.9 $\pm$ 6.7 \\
 & 200 & luminosity & False & 2M++ & cloned & 1.03 $\pm$ 0.05 & 165.2 $\pm$ 7.8 & 1.03 $\pm$ 0.05 & 148.4 $\pm$ 6.2 \\
\textbf{M} & \textbf{200} & \textbf{luminosity} & \textbf{True} & \textbf{2M++} & \textbf{cloned} & \textbf{1.03 $\pm$ 0.05} & \textbf{166.9 $\pm$ 7.7} & \textbf{1.04 $\pm$ 0.05} & \textbf{152.8 $\pm$ 5.7} \\
O & 200 & number & True & 2M++ & cloned & 0.97 $\pm$ 0.06 & 171.6 $\pm$ 7.6 & 0.98 $\pm$ 0.06 & 155.7 $\pm$ 5.7  \\
 & 200 & halo & False & 2M++ & uniform & 1.08 $\pm$ 0.06 & 171.0 $\pm$ 8.3 & 1.1 $\pm$ 0.06 & 154.1 $\pm$ 6.8 \\
 & 200 & luminosity & False & 2M++ & uniform & 1.06 $\pm$ 0.04 & 165.6 $\pm$ 7.7 & 1.06 $\pm$ 0.04 & 148.8 $\pm$ 6.0  \\
N & 200 & luminosity & True & 2M++ & uniform & 1.07 $\pm$ 0.04 & 167.0 $\pm$ 7.7 & 1.07 $\pm$ 0.05 & 153.1 $\pm$ 5.6  \\
P & 200 & number & True & 2M++ & uniform & 1.01 $\pm$ 0.05 & 172.0 $\pm$ 7.8 & 1.01 $\pm$ 0.05 & 156.3 $\pm$ 5.9 \\

\end{tabular}
\end{table*}

\begin{table*}\caption{Column and row separations have the same meaning as in Table \ref{table:Tab1}, except here we show our measurements of $\zeta^{\dagger} = \beta\sbr{norm,t}/f/\sigma\sbr{8,m}$ for the flux limited density tracers only.}
\label{table:Tab2}
\begin{tabular}{llllllllll}
& max R &  & flux & Applied& ZoA  &  SAG  &  SAG  &  SAGE  &  SAGE   \\
& [$\h$ Mpc] & tracer type & limited &  ZoA & Treatment &  $\zeta^\dagger$  &  $ \sigma_v$ &  $\zeta^\dagger$ & $ \sigma_v$ \\ \hline
 & 200 & luminosity & True & -- & n/a & 1.05 $\pm$ 0.03 & 161.1 $\pm$ 6.0 & 1.05 $\pm$ 0.03 & 146.7 $\pm$ 4.2 \\
& 200 & number & True & -- & n/a & 1.0 $\pm$ 0.06 & 164.4 $\pm$ 6.5 & 1.0 $\pm$ 0.06 & 148.3 $\pm$ 4.8  \\
 & 125 & luminosity & True & -- & n/a & 1.06 $\pm$ 0.06 & 159.6 $\pm$ 12.7 & 1.06 $\pm$ 0.05 & 145.4 $\pm$ 8.8 \\
 & 125 & number & True & -- & n/a & 1.0 $\pm$ 0.09 & 163.0 $\pm$ 12.4 & 1.02 $\pm$ 0.07 & 146.9 $\pm$ 8.7  \\
 \hline
 & 200 & luminosity & True & $10^\circ$ & cloned &  0.99 $\pm$ 0.04 & 167.9 $\pm$ 7.0 & 0.99 $\pm$ 0.04 & 153.9 $\pm$ 5.3 \\
 & 200 & number & True & $10^\circ$ & cloned & 0.94 $\pm$ 0.06 & 171.0 $\pm$ 7.4 & 0.94 $\pm$ 0.07 & 155.4 $\pm$ 5.9 \\
 & 200 & luminosity & True & $10^\circ$ & uniform & 1.08 $\pm$ 0.04 & 167.6 $\pm$ 6.9 & 1.08 $\pm$ 0.04 & 153.7 $\pm$ 5.0 \\
 & 200 & number & True & $10^\circ$ & uniform & 1.03 $\pm$ 0.05 & 171.6 $\pm$ 7.2 & 1.03 $\pm$ 0.05 & 156.0 $\pm$ 5.4 \\
 & 125 & luminosity & True & $5^\circ$ & cloned & 1.03 $\pm$ 0.06 & 162.7 $\pm$ 12.7 & 1.03 $\pm$ 0.05 & 148.5 $\pm$ 8.56  \\
 & 125 & number & True & $5^\circ$ & cloned & 0.98 $\pm$ 0.09 & 165.9 $\pm$ 12.5 & 0.99 $\pm$ 0.08 & 150.0 $\pm$ 8.3 \\
 & 125 & luminosity & True & $5^\circ$ & uniform &1.08 $\pm$ 0.07 & 163.6 $\pm$ 13.1 & 1.09 $\pm$ 0.06 & 149.4 $\pm$ 9.0   \\
 & 125 & number & True & $5^\circ$ & uniform & 1.03 $\pm$ 0.09 & 168.1 $\pm$ 13.3 & 1.04 $\pm$ 0.08 & 152.3 $\pm$ 9.4 \\
\hline
\textbf{Q} & \textbf{200} & \textbf{luminosity} & \textbf{True} & \textbf{2M++} & \textbf{cloned} & \textbf{1.04 $\pm$ 0.05} & \textbf{167.6 $\pm$ 7.7} & \textbf{1.04 $\pm$ 0.05} & \textbf{153.6 $\pm$ 5.9} \\
S & 200 & number & True & 2M++ & cloned & .98 $\pm$ 0.07 & 172.3 $\pm$ 8.0 & 0.98 $\pm$ 0.07 & 156.4 $\pm$ 6.0  \\
R & 200 & luminosity & True & 2M++ & uniform &1.08 $\pm$ 0.05 & 167.4 $\pm$ 7.7 & 1.08 $\pm$ 0.05 & 153.5 $\pm$ 5.7  \\
T & 200 & number & True & 2M++ & uniform & 1.02 $\pm$ 0.05 & 172.1 $\pm$ 8.0 & 1.02 $\pm$ 0.05 & 156.5 $\pm$ 6.1 \\
\end{tabular}
\end{table*}

%% file: main.bbl
\begin{thebibliography}{}
\makeatletter
\relax
\def\mn@urlcharsother{\let\do\@makeother \do\$\do\&\do\#\do\^\do\_\do\%\do\~}
\def\mn@doi{\begingroup\mn@urlcharsother \@ifnextchar [ {\mn@doi@}
  {\mn@doi@[]}}
\def\mn@doi@[#1]#2{\def\@tempa{#1}\ifx\@tempa\@empty \href
  {http://dx.doi.org/#2} {doi:#2}\else \href {http://dx.doi.org/#2} {#1}\fi
  \endgroup}
\def\mn@eprint#1#2{\mn@eprint@#1:#2::\@nil}
\def\mn@eprint@arXiv#1{\href {http://arxiv.org/abs/#1} {{\tt arXiv:#1}}}
\def\mn@eprint@dblp#1{\href {http://dblp.uni-trier.de/rec/bibtex/#1.xml}
  {dblp:#1}}
\def\mn@eprint@#1:#2:#3:#4\@nil{\def\@tempa {#1}\def\@tempb {#2}\def\@tempc
  {#3}\ifx \@tempc \@empty \let \@tempc \@tempb \let \@tempb \@tempa \fi \ifx
  \@tempb \@empty \def\@tempb {arXiv}\fi \@ifundefined
  {mn@eprint@\@tempb}{\@tempb:\@tempc}{\expandafter \expandafter \csname
  mn@eprint@\@tempb\endcsname \expandafter{\@tempc}}}

\bibitem[\protect\citeauthoryear{Aaronson, Huchra  \& Mould}{Aaronson
  et~al.}{1979}]{aaronson_infrared_1979}
Aaronson M.,  Huchra J.,   Mould J.,  1979, \mn@doi [\apj] {10.1086/156923},
  229, 1

\bibitem[\protect\citeauthoryear{Abazajian}{Abazajian}{2009}]{abazajian_seventh_2009}
Abazajian K.,  2009, \mn@doi [\aaps] {10.1088/0067-0049/182/2/543}, 182, 543

\bibitem[\protect\citeauthoryear{{Abbott} et~al.,}{{Abbott}
  et~al.}{2022}]{abbott_dark_2022}
{Abbott} T.~M.~C.,  et~al., 2022, \mn@doi [\prd] {10.1103/PhysRevD.105.023520},
  \href {https://ui.adsabs.harvard.edu/abs/2022PhRvD.105b3520A} {105, 023520}

\bibitem[\protect\citeauthoryear{Adams \& Blake}{Adams \&
  Blake}{2017}]{adams_improving_2017}
Adams C.,  Blake C.,  2017, \mn@doi [\mnras] {10.1093/mnras/stx1529}, 471, 839

\bibitem[\protect\citeauthoryear{Adams \& Blake}{Adams \&
  Blake}{2020}]{adams_joint_2020}
Adams C.,  Blake C.,  2020, \mn@doi [\mnras] {10.1093/mnras/staa845}, 494, 3275

\bibitem[\protect\citeauthoryear{{Alam} et~al.,}{{Alam}
  et~al.}{2021}]{alam_completed_2021}
{Alam} S.,  et~al., 2021, \mn@doi [\prd] {10.1103/PhysRevD.103.083533}, \href
  {https://ui.adsabs.harvard.edu/abs/2021PhRvD.103h3533A} {103, 083533}

\bibitem[\protect\citeauthoryear{{Amon} \& {Efstathiou}}{{Amon} \&
  {Efstathiou}}{2022}]{amon_nonlinear_2022}
{Amon} A.,  {Efstathiou} G.,  2022, \mn@doi [\mnras] {10.1093/mnras/stac2429},
  \href {https://ui.adsabs.harvard.edu/abs/2022MNRAS.516.5355A} {516, 5355}

\bibitem[\protect\citeauthoryear{{Asgari} et~al.,}{{Asgari}
  et~al.}{2021}]{asgari_kids-1000_2021}
{Asgari} M.,  et~al., 2021, \mn@doi [\aap] {10.1051/0004-6361/202039070}, \href
  {https://ui.adsabs.harvard.edu/abs/2021A&A...645A.104A} {645, A104}

\bibitem[\protect\citeauthoryear{Baker, Ferreira, Leonard  \& Motta}{Baker
  et~al.}{2014}]{baker_new_2014}
Baker T.,  Ferreira P.~G.,  Leonard C.~D.,   Motta M.,  2014, \mn@doi [\prd]
  {10.1103/PhysRevD.90.124030}, 90, 124030

\bibitem[\protect\citeauthoryear{Behroozi, Conroy  \& Wechsler}{Behroozi
  et~al.}{2010}]{behroozi_comprehensive_2010}
Behroozi P.~S.,  Conroy C.,   Wechsler R.~H.,  2010, \mn@doi [\apj]
  {10.1088/0004-637X/717/1/379}, 717, 379

\bibitem[\protect\citeauthoryear{Behroozi, Wechsler  \& Wu}{Behroozi
  et~al.}{2012}]{behroozi_rockstar_2012}
Behroozi P.~S.,  Wechsler R.~H.,   Wu H.-Y.,  2012, \mn@doi [\apj]
  {10.1088/0004-637X/762/2/109}, 762, 109

\bibitem[\protect\citeauthoryear{Behroozi, Wechsler, Wu, Busha, Klypin  \&
  Primack}{Behroozi et~al.}{2013}]{behroozi_gravitationally_2013}
Behroozi P.~S.,  Wechsler R.~H.,  Wu H.-Y.,  Busha M.~T.,  Klypin A.~A.,
  Primack J.~R.,  2013, \mn@doi [\apj] {10.1088/0004-637X/763/1/18}, 763, 18

\bibitem[\protect\citeauthoryear{Boruah, Hudson  \& Lavaux}{Boruah
  et~al.}{2020}]{boruah_cosmic_2020}
Boruah S.~S.,  Hudson M.~J.,   Lavaux G.,  2020, \mn@doi [\mnras]
  {10.1093/mnras/staa2485}, 498, 2703

\bibitem[\protect\citeauthoryear{{Brout} et~al.,}{{Brout}
  et~al.}{2022}]{brout_pantheon_2022}
{Brout} D.,  et~al., 2022, \mn@doi [\apj] {10.3847/1538-4357/ac8e04}, \href
  {https://ui.adsabs.harvard.edu/abs/2022ApJ...938..110B} {938, 110}

\bibitem[\protect\citeauthoryear{Carrick, Turnbull, Lavaux  \& Hudson}{Carrick
  et~al.}{2015}]{carrick_cosmological_2015}
Carrick J.,  Turnbull S.~J.,  Lavaux G.,   Hudson M.~J.,  2015, \mn@doi
  [\mnras] {10.1093/mnras/stv547}, 450, 317

\bibitem[\protect\citeauthoryear{Cora et~al.,}{Cora
  et~al.}{2018}]{cora_semi-analytic_2018}
Cora S.~A.,  et~al., 2018, \mn@doi [\mnras] {10.1093/mnras/sty1131}, 479, 2

\bibitem[\protect\citeauthoryear{Croton et~al.,}{Croton
  et~al.}{2016}]{croton_semi-analytic_2016}
Croton D.~J.,  et~al., 2016, \mn@doi [\aaps] {10.3847/0067-0049/222/2/22}, 222,
  22

\bibitem[\protect\citeauthoryear{{Dalal} et~al.,}{{Dalal}
  et~al.}{2023}]{dalal_hyper_2023}
{Dalal} R.,  et~al., 2023, \mn@doi [\prd] {10.1103/PhysRevD.108.123519}, \href
  {https://ui.adsabs.harvard.edu/abs/2023PhRvD.108l3519D} {108, 123519}

\bibitem[\protect\citeauthoryear{Davis \& Huchra}{Davis \&
  Huchra}{1982}]{davis_survey_1982}
Davis M.,  Huchra J.,  1982, \mn@doi [\apj] {10.1086/159751}, 254, 437

\bibitem[\protect\citeauthoryear{Davis, Nusser, Masters, Springob, Huchra  \&
  Lemson}{Davis et~al.}{2011}]{davis_local_2011}
Davis M.,  Nusser A.,  Masters K.~L.,  Springob C.,  Huchra J.~P.,   Lemson G.,
   2011, \mn@doi [\mnras] {10.1111/j.1365-2966.2011.18362.x}, 413, 2906

\bibitem[\protect\citeauthoryear{Dupuy, Courtois  \& Kubik}{Dupuy
  et~al.}{2019}]{dupuy_estimation_2019}
Dupuy A.,  Courtois H.~M.,   Kubik B.,  2019, \mn@doi [\mnras]
  {10.1093/mnras/stz901}, 486, 440

\bibitem[\protect\citeauthoryear{Gorski, Davis, Strauss, White  \&
  Yahil}{Gorski et~al.}{1989}]{gorski_cosmological_1989}
Gorski K.~M.,  Davis M.,  Strauss M.~A.,  White S. D.~M.,   Yahil A.,  1989,
  \mn@doi [\apj] {10.1086/167771}, 344, 1

\bibitem[\protect\citeauthoryear{{Hellwing}, {Schaller}, {Frenk}, {Theuns},
  {Schaye}, {Bower}  \& {Crain}}{{Hellwing}
  et~al.}{2016}]{hellwing_effect_2016}
{Hellwing} W.~A.,  {Schaller} M.,  {Frenk} C.~S.,  {Theuns} T.,  {Schaye} J.,
  {Bower} R.~G.,   {Crain} R.~A.,  2016, \mn@doi [\mnras]
  {10.1093/mnrasl/slw081}, \href
  {https://ui.adsabs.harvard.edu/abs/2016MNRAS.461L..11H} {461, L11}

\bibitem[\protect\citeauthoryear{Heymans et~al.,}{Heymans
  et~al.}{2013}]{heymans_cfhtlens_2013}
Heymans C.,  et~al., 2013, \mn@doi [\mnras] {10.1093/mnras/stt601}, 432, 2433

\bibitem[\protect\citeauthoryear{Heymans et~al.,}{Heymans
  et~al.}{2021}]{heymans_kids-1000_2021}
Heymans C.,  et~al., 2021, \mn@doi [\aap] {10.1051/0004-6361/202039063}, 646,
  A140

\bibitem[\protect\citeauthoryear{Hollinger \& Hudson}{Hollinger \&
  Hudson}{2021}]{hollinger_assessing_2021}
Hollinger A.~M.,  Hudson M.~J.,  2021, \mn@doi [\mnras]
  {10.1093/mnras/staa4039}, 502, 3723

\bibitem[\protect\citeauthoryear{Howlett}{Howlett}{2019}]{howlett_redshift-space_2019}
Howlett C.,  2019, \mn@doi [\mnras] {10.1093/mnras/stz1403}, 487, 5209

\bibitem[\protect\citeauthoryear{Howlett, Staveley-Smith  \& Blake}{Howlett
  et~al.}{2017a}]{howlett_cosmological_2017}
Howlett C.,  Staveley-Smith L.,   Blake C.,  2017a, \mn@doi [\mnras]
  {10.1093/mnras/stw2466}, 464, 2517

\bibitem[\protect\citeauthoryear{Howlett, Robotham, Lagos  \& Kim}{Howlett
  et~al.}{2017b}]{howlett_measuring_2017}
Howlett C.,  Robotham A. S.~G.,  Lagos C. D.~P.,   Kim A.~G.,  2017b, \mn@doi
  [\apj] {10.3847/1538-4357/aa88c8}, 847, 128

\bibitem[\protect\citeauthoryear{Huchra \& Geller}{Huchra \&
  Geller}{1982}]{huchra_groups_1982}
Huchra J.~P.,  Geller M.~J.,  1982, \mn@doi [\apj] {10.1086/160000}, 257, 423

\bibitem[\protect\citeauthoryear{Huchra et~al.,}{Huchra
  et~al.}{2012}]{huchra_2mass_2012}
Huchra J.~P.,  et~al., 2012, \mn@doi [\apjs] {10.1088/0067-0049/199/2/26}, 199,
  26

\bibitem[\protect\citeauthoryear{Hudson \& Turnbull}{Hudson \&
  Turnbull}{2012}]{hudson_growth_2012}
Hudson M.~J.,  Turnbull S.~J.,  2012, \mn@doi [\apjl]
  {10.1088/2041-8205/751/2/L30}, 751, L30

\bibitem[\protect\citeauthoryear{Hudson, Lucey, Smith, Schlegel  \&
  Davies}{Hudson et~al.}{2001}]{hudson_streaming_2001}
Hudson M.~J.,  Lucey J.~R.,  Smith R.~J.,  Schlegel D.~J.,   Davies R.~L.,
  2001, \mn@doi [\mnras] {10.1046/j.1365-8711.2001.04786.x}, 327, 265

\bibitem[\protect\citeauthoryear{Huterer, Shafer, Scolnic  \& Schmidt}{Huterer
  et~al.}{2017}]{huterer_testing_2017}
Huterer D.,  Shafer D.~L.,  Scolnic D.~M.,   Schmidt F.,  2017, \mn@doi [\jcap]
  {10.1088/1475-7516/2017/05/015}, 2017, 015

\bibitem[\protect\citeauthoryear{Johnson et~al.,}{Johnson
  et~al.}{2014}]{johnson_6df_2014}
Johnson A.,  et~al., 2014, \mn@doi [\mnras] {10.1093/mnras/stu1615}, 444, 3926

\bibitem[\protect\citeauthoryear{{Johnson}, {Blake}, {Dossett}, {Koda},
  {Parkinson}  \& {Joudaki}}{{Johnson} et~al.}{2016}]{johnson_searching_2016}
{Johnson} A.,  {Blake} C.,  {Dossett} J.,  {Koda} J.,  {Parkinson} D.,
  {Joudaki} S.,  2016, \mn@doi [\mnras] {10.1093/mnras/stw447}, \href
  {https://ui.adsabs.harvard.edu/abs/2016MNRAS.458.2725J} {458, 2725}

\bibitem[\protect\citeauthoryear{Jones et~al.,}{Jones
  et~al.}{2009}]{jones_6df_2009}
Jones D.~H.,  et~al., 2009, \mn@doi [\mnras]
  {10.1111/j.1365-2966.2009.15338.x}, 399, 683

\bibitem[\protect\citeauthoryear{Juszkiewicz, Feldman, Fry  \&
  Jaffe}{Juszkiewicz et~al.}{2010}]{juszkiewicz_weakly_2010}
Juszkiewicz R.,  Feldman H.~A.,  Fry J.,   Jaffe A.~H.,  2010, \mn@doi [\jcap]
  {10.1088/1475-7516/2010/02/021}, 2010, 021

\bibitem[\protect\citeauthoryear{Kenworthy et~al.,}{Kenworthy
  et~al.}{2022}]{kenworthy_measurements_2022}
Kenworthy W.~D.,  et~al., 2022, \mn@doi [\apj] {10.3847/1538-4357/ac80bd}, 935,
  83

\bibitem[\protect\citeauthoryear{Klypin, Yepes, Gottlöber, Prada  \&
  Heß}{Klypin et~al.}{2016}]{klypin_multidark_2016}
Klypin A.,  Yepes G.,  Gottlöber S.,  Prada F.,   Heß S.,  2016, \mn@doi
  [\mnras] {10.1093/mnras/stw248}, 457, 4340

\bibitem[\protect\citeauthoryear{Knebe et~al.,}{Knebe
  et~al.}{2018}]{knebe_multidark-galaxies_2018}
Knebe A.,  et~al., 2018, \mn@doi [\mnras] {10.1093/mnras/stx2662}, 474, 5206

\bibitem[\protect\citeauthoryear{Koda et~al.,}{Koda
  et~al.}{2014}]{koda_are_2014}
Koda J.,  et~al., 2014, \mn@doi [\mnras] {10.1093/mnras/stu1610}, 445, 4267

\bibitem[\protect\citeauthoryear{Kravtsov, Berlind, Wechsler, Klypin,
  Gottlöber, Allgood  \& Primack}{Kravtsov et~al.}{2004}]{kravtsov_dark_2004}
Kravtsov A.~V.,  Berlind A.~A.,  Wechsler R.~H.,  Klypin A.~A.,  Gottlöber S.,
   Allgood B.,   Primack J.~R.,  2004, \mn@doi [\apj] {10.1086/420959}, 609, 35

\bibitem[\protect\citeauthoryear{Lavaux \& Hudson}{Lavaux \&
  Hudson}{2011}]{lavaux_2m_2011}
Lavaux G.,  Hudson M.~J.,  2011, \mn@doi [\mnras]
  {10.1111/j.1365-2966.2011.19233.x}, 416, 2840

\bibitem[\protect\citeauthoryear{Lilow \& Nusser}{Lilow \&
  Nusser}{2021}]{lilow_constrained_2021}
Lilow R.,  Nusser A.,  2021, \mn@doi [\mnras] {10.1093/mnras/stab2009}, 507,
  1557

\bibitem[\protect\citeauthoryear{Linder}{Linder}{2005}]{linder_cosmic_2005}
Linder E.~V.,  2005, \prd, 72, 043529

\bibitem[\protect\citeauthoryear{Lynden-Bell, Faber, Burstein, Davies,
  Dressler, Terlevich  \& Wegner}{Lynden-Bell
  et~al.}{1988}]{lynden-bell_photometry_1988}
Lynden-Bell D.,  Faber S.~M.,  Burstein D.,  Davies R.~L.,  Dressler A.,
  Terlevich R.~J.,   Wegner G.,  1988, \mn@doi [\apj] {10.1086/166066}, 326, 19

\bibitem[\protect\citeauthoryear{Lynden-Bell, Lahav  \& Burstein}{Lynden-Bell
  et~al.}{1989}]{lynden-bell_cosmological_1989}
Lynden-Bell D.,  Lahav O.,   Burstein D.,  1989, \mn@doi [\mnras]
  {10.1093/mnras/241.2.325}, 241, 325

\bibitem[\protect\citeauthoryear{{Madhavacheril} et~al.,}{{Madhavacheril}
  et~al.}{2024}]{madhavacheril_atacama_2024}
{Madhavacheril} M.~S.,  et~al., 2024, \mn@doi [\apj]
  {10.3847/1538-4357/acff5f}, \href
  {https://ui.adsabs.harvard.edu/abs/2024ApJ...962..113M} {962, 113}

\bibitem[\protect\citeauthoryear{Marinoni \& Hudson}{Marinoni \&
  Hudson}{2002}]{marinoni_mass--light_2002}
Marinoni C.,  Hudson M.~J.,  2002, \mn@doi [\apj] {10.1086/339319}, 569, 101

\bibitem[\protect\citeauthoryear{{Mead}, {Heymans}, {Lombriser}, {Peacock},
  {Steele}  \& {Winther}}{{Mead} et~al.}{2016}]{mead_accurate_2016}
{Mead} A.~J.,  {Heymans} C.,  {Lombriser} L.,  {Peacock} J.~A.,  {Steele}
  O.~I.,   {Winther} H.~A.,  2016, \mn@doi [\mnras] {10.1093/mnras/stw681},
  \href {https://ui.adsabs.harvard.edu/abs/2016MNRAS.459.1468M} {459, 1468}

\bibitem[\protect\citeauthoryear{{Mead}, {Brieden}, {Tr{\"o}ster}  \&
  {Heymans}}{{Mead} et~al.}{2021}]{mead_hmcode_2021}
{Mead} A.~J.,  {Brieden} S.,  {Tr{\"o}ster} T.,   {Heymans} C.,  2021, \mn@doi
  [\mnras] {10.1093/mnras/stab082}, \href
  {https://ui.adsabs.harvard.edu/abs/2021MNRAS.502.1401M} {502, 1401}

\bibitem[\protect\citeauthoryear{Moster, Somerville, Maulbetsch, Bosch,
  Macciò, Naab  \& Oser}{Moster et~al.}{2010}]{moster_constraints_2010}
Moster B.~P.,  Somerville R.~S.,  Maulbetsch C.,  Bosch F. C. v.~d.,  Macciò
  A.~V.,  Naab T.,   Oser L.,  2010, \mn@doi [\apj]
  {10.1088/0004-637X/710/2/903}, 710, 903

\bibitem[\protect\citeauthoryear{Neill, Hudson  \& Conley}{Neill
  et~al.}{2007}]{neill_peculiar_2007}
Neill J.~D.,  Hudson M.~J.,   Conley A.,  2007, \mn@doi [\apj]
  {10.1086/518808}, 661, L123

\bibitem[\protect\citeauthoryear{{Nguyen}, {Huterer}  \& {Wen}}{{Nguyen}
  et~al.}{2023}]{nguyen_evidence_2023}
{Nguyen} N.-M.,  {Huterer} D.,   {Wen} Y.,  2023, \mn@doi [\prl]
  {10.1103/PhysRevLett.131.111001}, \href
  {https://ui.adsabs.harvard.edu/abs/2023PhRvL.131k1001N} {131, 111001}

\bibitem[\protect\citeauthoryear{Nusser}{Nusser}{2017}]{nusser_velocity-density_2017}
Nusser A.,  2017, \mn@doi [\mnras] {10.1093/mnras/stx1225}, 470, 445

\bibitem[\protect\citeauthoryear{{Pan} et~al.,}{{Pan}
  et~al.}{2023}]{pan_measurement_2023}
{Pan} Z.,  et~al., 2023, \mn@doi [\prd] {10.1103/PhysRevD.108.122005}, \href
  {https://ui.adsabs.harvard.edu/abs/2023PhRvD.108l2005P} {108, 122005}

\bibitem[\protect\citeauthoryear{Park}{Park}{2000}]{park_cosmic_2000}
Park C.,  2000, \mn@doi [\mnras] {10.1111/j.1365-8711.2000.03886.x}, 319, 573

\bibitem[\protect\citeauthoryear{Park \& Park}{Park \&
  Park}{2006}]{park_power_2006}
Park C.-G.,  Park C.,  2006, \mn@doi [\apj] {10.1086/498258}, 637, 1

\bibitem[\protect\citeauthoryear{Peterson et~al.,}{Peterson
  et~al.}{2022}]{peterson_pantheon_2022}
Peterson E.~R.,  et~al., 2022, \mn@doi [\apj] {10.3847/1538-4357/ac4698}, 938,
  112

\bibitem[\protect\citeauthoryear{Pike \& Hudson}{Pike \&
  Hudson}{2005}]{pike_cosmological_2005}
Pike R.,  Hudson M.~J.,  2005, \apj, 635, 11

\bibitem[\protect\citeauthoryear{{Planck Collaboration} et~al.,}{{Planck
  Collaboration} et~al.}{2020a}]{planck_collaboration_planck_2020}
{Planck Collaboration} et~al., 2020a, \mn@doi [\aap]
  {10.1051/0004-6361/201833880}, 641, A1

\bibitem[\protect\citeauthoryear{{Planck Collaboration} et~al.,}{{Planck
  Collaboration} et~al.}{2020b}]{planck_collaboration_planck_2020-1}
{Planck Collaboration} et~al., 2020b, \mn@doi [\aap]
  {10.1051/0004-6361/201833910}, 641, A6

\bibitem[\protect\citeauthoryear{{Planck Collaboration} et~al.,}{{Planck
  Collaboration} et~al.}{2020c}]{planck_collaboration_planck_2020-7}
{Planck Collaboration} et~al., 2020c, \mn@doi [\aap]
  {10.1051/0004-6361/201833886}, \href
  {https://ui.adsabs.harvard.edu/abs/2020A&A...641A...8P} {641, A8}

\bibitem[\protect\citeauthoryear{Qin, Howlett  \& Staveley-Smith}{Qin
  et~al.}{2019}]{qin_redshift-space_2019}
Qin F.,  Howlett C.,   Staveley-Smith L.,  2019, \mn@doi [\mnras]
  {10.1093/mnras/stz1576}, 487, 5235

\bibitem[\protect\citeauthoryear{Riess et~al.,}{Riess
  et~al.}{2016}]{riess_24_2016}
Riess A.~G.,  et~al., 2016, \mn@doi [\apj] {10.3847/0004-637X/826/1/56}, 826,
  56

\bibitem[\protect\citeauthoryear{Riess et~al.,}{Riess
  et~al.}{2018}]{riess_new_2018}
Riess A.~G.,  et~al., 2018, \mn@doi [\apj] {10.3847/1538-4357/aaadb7}, 855, 136

\bibitem[\protect\citeauthoryear{Riess et~al.,}{Riess
  et~al.}{2022}]{riess_comprehensive_2022}
Riess A.~G.,  et~al., 2022, \mn@doi [\apjl] {10.3847/2041-8213/ac5c5b}, 934, L7

\bibitem[\protect\citeauthoryear{Said, Colless, Magoulas, Lucey  \&
  Hudson}{Said et~al.}{2020}]{said_joint_2020}
Said K.,  Colless M.,  Magoulas C.,  Lucey J.~R.,   Hudson M.~J.,  2020,
  \mn@doi [\mnras] {10.1093/mnras/staa2032}, 497, 1275

\bibitem[\protect\citeauthoryear{Sandage \& Tammann}{Sandage \&
  Tammann}{1975}]{sandage_steps_1975}
Sandage A.,  Tammann G.~A.,  1975, \mn@doi [\apj] {10.1086/153413}, 196, 313

\bibitem[\protect\citeauthoryear{{Schaller} et~al.,}{{Schaller}
  et~al.}{2015}]{schaller_baryon_2015}
{Schaller} M.,  et~al., 2015, \mn@doi [\mnras] {10.1093/mnras/stv1067}, \href
  {https://ui.adsabs.harvard.edu/abs/2015MNRAS.451.1247S} {451, 1247}

\bibitem[\protect\citeauthoryear{Schechter}{Schechter}{1976}]{schechter_analytic_1976}
Schechter P.,  1976, \mn@doi [\apj] {10.1086/154079}, 203, 297

\bibitem[\protect\citeauthoryear{Scolnic et~al.,}{Scolnic
  et~al.}{2018}]{scolnic_complete_2018}
Scolnic D.~M.,  et~al., 2018, \mn@doi [\apj] {10.3847/1538-4357/aab9bb}, 859,
  101

\bibitem[\protect\citeauthoryear{Smith et~al.,}{Smith
  et~al.}{2003}]{smith_stable_2003}
Smith R.~E.,  et~al., 2003, \mn@doi [\mnras]
  {10.1046/j.1365-8711.2003.06503.x}, 341, 1311

\bibitem[\protect\citeauthoryear{{Strauss} \& {Willick}}{{Strauss} \&
  {Willick}}{1995}]{strauss_density_1995}
{Strauss} M.~A.,  {Willick} J.~A.,  1995, \mn@doi [\physrep]
  {10.1016/0370-1573(95)00013-7}, \href
  {https://ui.adsabs.harvard.edu/abs/1995PhR...261..271S} {261, 271}

\bibitem[\protect\citeauthoryear{Takahashi, Sato, Nishimichi, Taruya  \&
  Oguri}{Takahashi et~al.}{2012}]{takahashi_revising_2012}
Takahashi R.,  Sato M.,  Nishimichi T.,  Taruya A.,   Oguri M.,  2012, \mn@doi
  [\apj] {10.1088/0004-637X/761/2/152}, 761, 152

\bibitem[\protect\citeauthoryear{Tasitsiomi, Kravtsov, Wechsler  \&
  Primack}{Tasitsiomi et~al.}{2004}]{tasitsiomi_modeling_2004}
Tasitsiomi A.,  Kravtsov A.~V.,  Wechsler R.~H.,   Primack J.~R.,  2004,
  \mn@doi [\apj] {10.1086/423784}, 614, 533

\bibitem[\protect\citeauthoryear{Tully et~al.,}{Tully
  et~al.}{2023}]{tully_cosmicflows-4_2023}
Tully R.~B.,  et~al., 2023, \mn@doi [\apj] {10.3847/1538-4357/ac94d8}, 944, 94

\bibitem[\protect\citeauthoryear{Turnbull, Hudson, Feldman, Hicken, Kirshner
  \& Watkins}{Turnbull et~al.}{2012}]{turnbull_cosmic_2012}
Turnbull S.~J.,  Hudson M.~J.,  Feldman H.~A.,  Hicken M.,  Kirshner R.~P.,
  Watkins R.,  2012, \mn@doi [\mnras] {10.1111/j.1365-2966.2011.20050.x}, 420,
  447

\bibitem[\protect\citeauthoryear{Turner, Blake  \& Ruggeri}{Turner
  et~al.}{2021}]{turner_improving_2021}
Turner R.~J.,  Blake C.,   Ruggeri R.,  2021, \mn@doi [\mnras]
  {10.1093/mnras/stab212}, 502, 2087

\bibitem[\protect\citeauthoryear{\VAN{Daalen}{Van}{van}~Daalen, {Schaye},
  {McCarthy}, {Booth}  \& {Dalla Vecchia}}{\VAN{Daalen}{Van}{van}~Daalen
  et~al.}{2014}]{vaandaalen_impact_2014}
\VAN{Daalen}{Van}{van}~Daalen M.~P.,  {Schaye} J.,  {McCarthy} I.~G.,  {Booth}
  C.~M.,   {Dalla Vecchia} C.,  2014, \mn@doi [\mnras] {10.1093/mnras/stu482},
  \href {https://ui.adsabs.harvard.edu/abs/2014MNRAS.440.2997V} {440, 2997}

\bibitem[\protect\citeauthoryear{\VAN{Vaucouleurs}{De}{de}~Vaucouleurs \&
  Bollinger}{\VAN{Vaucouleurs}{De}{de}~Vaucouleurs \&
  Bollinger}{1979}]{de_vaucouleurs_extragalactic_1979}
\VAN{Vaucouleurs}{De}{de}~Vaucouleurs G.,  Bollinger G.,  1979, \mn@doi [\apj]
  {10.1086/157405}, 233, 433

\bibitem[\protect\citeauthoryear{{Velliscig} et~al.,}{{Velliscig}
  et~al.}{2015}]{velliscig_intrinsic_2015}
{Velliscig} M.,  et~al., 2015, \mn@doi [\mnras] {10.1093/mnras/stv2198}, \href
  {https://ui.adsabs.harvard.edu/abs/2015MNRAS.454.3328V} {454, 3328}

\bibitem[\protect\citeauthoryear{Westover}{Westover}{2007}]{westover_galaxy_2007}
Westover M.,  2007, PhD thesis, \url
  {https://ui.adsabs.harvard.edu/abs/2007PhDT.........3W}

\bibitem[\protect\citeauthoryear{Yahil, Strauss, Davis  \& Huchra}{Yahil
  et~al.}{1991}]{yahil_redshift_1991}
Yahil A.,  Strauss M.~A.,  Davis M.,   Huchra J.~P.,  1991, \mn@doi [\apj]
  {10.1086/169985}, 372, 380

\makeatother
\end{thebibliography}
